\documentclass{eslart}
\usepackage{amssymb}
\begin{document}
\input psfig.sty

\hoffset=0pt
\voffset=0pt
\hyphenation{cal-o-rim-e-ter}
\hyphenation{ped-es-tal}

\begin{frontmatter}

\title{\bf Measurement of stopping beam distributions\\
in the PIBETA detector}

\author[UVa]{E.~Frle\v z\thanksref{author}},
\thanks[author]{Corresponding author. Tel: +1--434--924--6786; 
fax: +1--434--924--4576. \\ {\sl E--mail address}: frlez@virginia. edu (E. Frle\v{z}).\hfill}
\author[UVa]{M.~Bychkov},
\author[UVa,VCU]{W.~Li}, and
\author[UVa]{D.~Po\v cani\'c}
\address[UVa]{Department of Physics, University of Virginia, 
Charlottesville, VA~22904-4714}
\address[VCU]{Department of Radiation Oncology, Virginia Commonwealth University,
Richmond, VA 23298-0058}

\begin{abstract}
Precise calculation of the geometrical acceptance of a large solid
angle detector with an integrated stopping target relies on
precise knowledge of the beam geometry. We describe four alternative 
methods that we used to measure the beam stopping distributions in the PIBETA 
detector active target: (i) light response of segmented target elements
to incident beam particles, (ii) back-tracking of charged particles from 
$\pi^+$ and $\mu^+$ decays using multi-wire proportional chambers, (iii) volume distribution of the 
Dalitz decay ($\pi^0\to \gamma\e^+e^-$) event vertices, and (iv) the opening angle
distribution of two $\pi^0$ photons originating from the beta decay of $\pi^+$ at rest. 
We demonstrate consistent results obtained by these four independent approaches and show 
how particular beam stopping distributions affect the detector's geometrical acceptance.
\par\noindent\hbox{\ }\par\noindent
\par\noindent\hbox{\ }\par\noindent

\end{abstract}
\end{frontmatter}
\vfill\eject

\section{Introduction}\label{sec:int}
\medskip
The PIBETA project at the Paul Scherrer Institute (PSI) is a program of 
measurements which aims to make a precise determination of 
the {$\pi^{+}\to\pi^{0}e^{+}\nu_e$\ } decay rate, initially with 
an uncertainty of $\sim$~0.5\%~\cite{Poc91,Poc04}. The small branching ratio 
($\simeq 1 \times 10^{-8}$) for the pion beta ($\pi\beta$) decay and 
the required high measurement precision impose stringent requirements 
on the experimental apparatus. 

The experimental signature of a {$\pi^{+}\to\pi^{0}e^{+}\nu_e$\/ event is 
determined by the prompt decay $\pi^0 \to \gamma\gamma$. The detector must be able 
to handle high event rates and cover a large solid angle with high efficiency for 
$\pi^0$ detection.  Efficient hardware suppression of background events requires 
good energy and timing resolution.  At the same time the system must operate 
with low systematic uncertainties and be subject to accurate calibration.

We have chosen to detect $\pi\beta$ decays at rest, and to use $\pi^+\to e^+\nu_e$ 
($\pi_{e2}$) decays for normalization of the absolute branching ratio. 
Consequently, our apparatus has the following main components~\cite{Frl03}:

\begin{itemize}
\item[(1)] a forward beam counter (BC), an active degrader (AD) and a segmented 
active target (AT) to stop the beam pions;
\item[(2)] two concentric cylindrical multi-wire proportional chambers 
(MWPC$_1$ and MWPC$_2$) surrounding the active target for charged particle tracking;
\item[(3)] a segmented fast veto scintillation counter (PV) surrounding the MWPC's, 
for the particle identification;
\item[(4)] a high resolution, 240-module, pure CsI segmented fast shower calorimeter
surrounding the active target and tracking detectors in a near-spherical
geometry;
\item[(5)] cosmic ray veto counters (CV) around the entire apparatus. 
\end{itemize}

The $\pi\beta$ branching ratio {$B_{\pi\beta}$ is evaluated from the following expression:
\begin{equation}
   B_{\pi\beta} = 
   {f_{\pi_{e2}}\cdot B_{\pi_{e2}} \over B_{\pi^0\to\gamma\gamma}} \cdot
   {N_{\pi\beta} \over N_{\pi_{e2}}}\cdot
   {A_{\pi_{e2}} \over A_{\pi\beta}} \cdot {\epsilon_{\pi_{e2}} \over \epsilon_{\pi\beta}},
\label{eq:br}
\end{equation} 
where $B_{\pi\beta}$, $B_{\pi_{e2}}$ and $B_{\pi^0\to\gamma\gamma}$
are the branching ratios for the $\pi\beta$, $\pi^+\to e^+\nu_e$ and 
$\pi^0\to\gamma\gamma$ processes, respectively, $N_{\pi\beta}$ and $N_{\pi_{e2}}$
represent the detected numbers of good $\pi\beta$ and $\pi^+\to e^+\nu_e$ events, 
$A_{\pi\beta}$ and $A_{\pi_{e2}}$ are the detector acceptances, $\epsilon_{\pi\beta}$ and 
$\epsilon_{\pi_{e2}}$ are the detector efficiencies, and $f_{\pi_{e2}}$ is
the prescaling factor for the $\pi^+\to e^+\nu_e$ trigger.

Due to the similarities between the two classes of events, the acceptances
$A_{\pi\beta}$ and $A_{\pi_{e2}}$ are nearly equal.  At the same time, both are
functions of
\begin{itemize}
\item[(1)] the geometrical solid angle covered by the calorimeter;
\item[(2)] geometry of a beam stopping distribution;
\item[(3)] energy line-shapes of photon- and positron-induced 
showers in the calorimeter;
\item[(4)] software cuts imposed on the measured event variables.
\end{itemize}

The main source of uncertainty of geometrical nature in the absolute acceptance 
ratio $A_{\pi_{e2}}/A_{\pi\beta}$ has to do with the position and spatial
spreading of the pion beam stopping distribution.  Due to the cylindrical
symmetry of our detector system the uncertainties can be broken down into
lateral ($x$--$y$ coordinates perpendicular to the beam direction) and axial 
uncertainties ($z$ coordinate in the direction of the beam axis). Of the two, 
the lateral uncertainties produce a larger effect on the acceptances than the axial
uncertainties. Also, since the $\pi\beta$ events require symmetrical coverage for 
the two photons emitted in the $\pi^0$ decay (nearly at rest), $A_{\pi\beta}$ is 
affected more than $A_{\pi_{e2}}$.

In this report we analyze the PIBETA data we have collected in order to monitor 
the centroid and shape of the pion beam stopping distribution. Our goal is to examine how 
the uncertainty in the deduced beam geometry affects the ${A_{\pi_{e2}}/A_{\pi\beta}}$ 
ratio. In Section~\ref{sec:beam} we specify the design and layout of the experimental beam line.
Sections~\ref{sec:assem}~and~\ref{sec:pos} address the related issue of geometrical uncertainties
associated with the assembly of the modular shower calorimeter.
In Section~\ref{sec:tgt} we describe the PIBETA detector's central region, most importantly 
the segmented active target. Section~\ref{sec:adc} lays out the method and specifies the accuracy 
with which the lateral distribution of stopping pions (and contaminating positrons) 
can be deduced from their respective calibrated pulse height spectra in the active target segments. 
Section~\ref{sec:tomo} introduces an alternative way of determining the beam position 
and shape that relies on the back-tracking of charged particles using MWPC hits. 
In Section~\ref{sec:dal} the beam stopping distribution is deduced from the reconstructed
vertices of the Dalitz decay events $\pi^0\to \gamma\e^+e^-$ which originate from 
the pion beta decays at rest.
Influence of the different beam distributions on 
(i) the absolute acceptance of the detector, and (ii) measured $\pi^0\to\gamma\gamma$ 
opening angle distributions, is calculated in a Monte Carlo (MC) detector simulation 
and presented in Section~\ref{sec:ang}.
Finally, the mutual consistency of the beam profile reconstructions by
these four independent methods is summarized in the Conclusion.

\bigskip
\section{PIBETA detector beam line}\label{sec:beam}
\medskip

The measurements are performed in the $\pi E1$ channel at PSI~\cite{Psi94}. 
The beam line is operated in the high intensity, low momentum resolution 
mode. A 114 MeV/c $\pi^+$ beam tune is developed with momentum spread 
$\Delta p/p\le 1.2\,$\% and maximum nominal $\pi^+$ beam intensity $I_\pi$\/ 
of $\simeq 1.2\cdot 10^6\,\pi^+$'s per second.

The spatial spread of the $\pi^+$ beam is restricted by a lead collimator with 
a 7$\,$mm pin-hole located 3.985$\,$m upstream of the detector center. The beam 
particles are first registered in a 3$\,$mm thick plastic scintillator placed 
directly behind the collimator. The pions are subsequently slowed in 
a 40$\,$mm long active plastic degrader and stopped in an active plastic target that 
is positioned in the center of the PIBETA detector. The beam line properties have 
been computed with TRANSPORT~\cite{Bro73} and TURTLE~\cite{Bro74} codes. 
The TURTLE momentum spectrum of $\pi^+$'s incident on the front face of 
the degrader counter is shown in Fig.~\ref{fig:beam_p}.

This specific choice of the incident beam momentum allows for the optimum 
pion/muon/positron separation by the time-of-flight (TOF) method. 
The absolute TOF's measured between the beam counter and the degrader/target 
detectors for the beam positrons, muons, and pions are 12.7$\,$ns, 17.4 $\,$ns, 
and 20.2$\,$ns, respectively.

The beam particle triggers are defined by a four-fold coincidence between 
(1) forward beam counter BC, (2) active degrader AD, (3) active target AT, and 
(4) the accelerator rf signal. The energy depositions of minimum-ionizing 
positrons in BC, AD, and AT counters are 0.6$\,$MeV, 7.2$\,$MeV, and 9.0$\,$MeV, 
respectively. The corresponding energy deposition values for 114$\,$MeV/c pions 
(with kinetic energy 40$\,$MeV) are 0.7$\,$MeV, 13.0$\,$MeV, and 28.0$\,$MeV, 
respectively. By appropriately adjusting the discriminator thresholds and 
time delays of the logic signals that make up the quad coincidence, the ``$\pi$-in-beam'' 
and ``$e$-in-beam'' triggers, which preferentially select beam pions and positrons, 
respectively, are set up. A digital oscilloscope snap-shot of four trigger-defining signals 
at the input of the ``$\pi$-in-beam'' coincidence unit is shown in Fig.~\ref{fig:pistop}.

The $e^+$ and $\mu^+$ beam contaminations in the ``$\pi$-in-beam'' trigger measured 
in the TOF spectra are small, $\simeq$\,0.4$\,$\% and $\simeq$\,0.2$\,$\%, respectively. 
\bigskip

\section{Design and Assembly of the CsI calorimeter sphere}\label{sec:assem}
\medskip

The design and assembly of the CsI calorimeter directly influence the precision
with which the geometry of the beam and the decay particle tracks can be determined.

The spherically shaped pure CsI calorimeter is mounted inside a forged
spherical steel shell which itself is supported by a steel frame. 
The shell has two large axial openings of 550\,mm diameter 
for beam entrance and exit and 220 smaller holes distributed over its surface 
and aligned with the axes of the individual CsI crystals. The smaller holes were made for 
access to the CsI photomultiplier tubes (PMT's). The beam openings were used to hold cast iron plugs 
with truncated pyramid shapes inside the shell to provide contact surfaces for 20
outlying shower veto crystals, respectively. These two plug plates press 
the stacked crystals together and hold them in place. The plugs have 
a cylindrical bore-hole (diameter 270\,mm) for the beam which also allows access 
to the interior of the CsI sphere, e.g. the wire chambers, the target counters, etc.

Assembly of the calorimeter, weighing approximately  1,500\,kg,
proceeded as follows: 

The empty spherical shell with its axis turned vertical,
supported by its steel frame, was equipped with a single
plug insert from below. The beam hole of this plug was used to hold a micrometer gauge
with a touch-ball in place. This device allowed the measurement of
a crystals' front surface distance to the center of the sphere with 100\,$\mu$m accuracy.
Afterwards, the whole frame was turned with the plug at the bottom.
Through the other beam hole, on top now and not yet closed with its plug, all
crystals already equipped with their PMT's were moved into the interior of the spherical
shell. First came 10 pieces of trapezohedron-shaped veto crystals which were
positioned on the flat surfaces of the upstanding pentagonal pyramid of the bottom
plug. The correct distance to the center of the sphere was adjusted using
threaded bolts mounted on the plug and supporting the crystals' rear end. 

After all 10 bottom veto crystals were in place, the next crystal layer
was stacked on top of the veto layer. Threaded tubes with bore-holes large enough
to house a PMT and its mu-metal shielding were installed through the smaller holes 
in the shell, which are aligned with the crystals' axes. The front end of those tubes
pressed against the rear surface of the crystals and thereby allowed for a
precise adjustment of their distances to the detector center point, as measured 
with the touch-ball measurement device.
Then the high voltage divider base was put through the tube and
connected with the PMT. Bases for the veto crystals had to be mounted somewhat 
differently as there are no holes available for them on the shell due to the close
proximity of the axial plugs. The closest hole, later employed for a crystal on
the next layer, was used to move the veto crystals base into the sphere, then sidewards
until connection the the veto's PMT was possible. Small supporting structures
were mounted to press the base gently against the PMT to prevent loosening
due to gravity's pull for vertically mounted crystals. 

When the sphere was half filled with crystals the procedure had to be
changed somewhat as the upper surface of completed layers became steeper and
steeper. In order to prevent the crystals from sliding down towards the sphere's
center, the measurement device at the center was replaced by a pin-holder with a
pin for each crystal still to be inserted and already adjusted in length to give
the correct distance of 260\,mm to the center.

After the last layer of 10 veto crystals was filled in, the second
pentagonally shaped plug was mounted on top and pressed against the crystal ball underneath. 
Calculations made prior to the assembly told us that pressing the plug
by 10\,mm against the crystal ball would be needed in order to pack all crystals tightly. 
This prediction turned out to be exactly fulfilled, giving us
confidence in the correct positioning of all modules. The completed setup was turned around
several times and watched for stability before the pin-holder at the center 
was disassembled through the beam entry/exit holes.

The detector geometry and location has been measured by the PSI surveying group using 
a theodolite. The upstream and downstream detector side are levelled, with the center 
$1506.0\pm 0.3\,$mm above the floor, the value corresponding to the nominal beam 
height in the $\pi$E1 channel. The measurement of the detector center was 
reproducible and the alignment of the target axis along the detector 
longitudinal axis is confirmed with 0.3$\,$mm uncertainty.

\bigskip
\section{Alignment of the MWPC's and the CsI calorimeter}\label{sec:pos}
\medskip

Two out of the four reconstruction methods described in this paper depend on 
the information provided by two independent detectors, namely the readings of 
the MWPC's and the signals from the CsI calorimeter. Great care was taken in 
the proper assembly and positioning of these two detectors as already described in
the previous section. During data acquisition, however, 
wire chambers required several maintenance sessions and they were once
removed from the entire assembly and then reinserted. This could 
have introduced different displacements between the wire chamber centers 
and the center of the CsI sphere. Such a displacement could significantly 
affect both the reconstruction of the beam's stopping distribution 
as well as the measured $\pi^{0} \to \gamma\gamma$ opening angle distribution. 
The following section describes how we ascertained the relative
offsets between the centers of the calorimeter and the wire chamber coordinate systems.

On a single event basis it is not possible to find the exact initial
point of interaction between a positron/photon and the face of 
the calorimeter relying only on the calorimeter signals. 
We can, however, determine whether a particular CsI detector was
close to a given track. Assuming that in such a case most of the energy
was deposited in the intersecting crystal, we can adopt the following procedure. 
We accept events with a single charged particle 
(e.g. $\mu \rightarrow e\nu_{e}\overline{\nu}_{\mu}$ positron)
and reconstruct the MWPC track and its intercept with the calorimeter. 
We then scan all the energies deposited in the calorimeter modules 
and find the maximum value. Calculated values of the intercept
coordinates corresponding to the crystal with most energy
deposited can be histogrammed. The outcome of this procedure is a set of individual
crystal maps, e.g., two-dimensional plots (polar angle $\theta$ versus azimuthal
angle $\phi$) of a given crystal
``illumination'' as seen by the wire chambers. Knowing the geometry of the
crystal we can predict how polar angle $\theta$ and azimuthal angle $\phi$ 
projected histograms should look and then compare them to the experimental data.

Since the inner radius of the calorimeter sphere is fixed at 260\,mm we have
effectively reduced our problem to a two-dimensional one. Namely, everything
depends only on the spherical coordinates $\theta$ and $\phi$ of the crystals. We
know that Michel events are uniformly distributed in solid angle.  Therefore,
if in $\theta$--$\phi$ space our crystals were rectangular, then the
probability density of hitting one particular crystal (for, say, the $\theta$
projection) would be ${dA}/{d\theta}$ (where $dA$ is infinitesimal
area of the $\theta$--$\phi$ projection) and therefore would be a
constant. In $\theta$--$\phi$ space our crystals do not have rectangular 
shapes and thus probability density is not a constant.  

Knowledge of the exact geometry of the crystals allows us to calculate the 
theoretical ${dA}/{d\theta}$ analytically. The complete derivation is presented
in Ref.~\cite{max}. Our $\theta$--$\phi$ plots are additionally 
smeared by the chamber resolution. Mathematically, it means that the 
${dA}/{d\theta}$ function is convoluted by some resolution function. 
In Sec.~\ref{sec:mwpc} we show that appropriate MWPC response is a Gaussian 
convolution and therefore the trial function is

\begin{equation}
f(m)=\int_{-\infty}^{+\infty} \frac{dA(\theta)}{d\theta}\cdot\Gamma\cdot
\exp{\left[ -\frac{(\theta-m)^{2}}{\sigma^{2}}\right]}d\theta,
\end{equation}
where $\Gamma$, $m$ and $\sigma$ are the parameters of the Gaussian
distribution, determined in a fit.

The exact position of the crystal enters the fitting function through the 
analytical expression for ${dA}/{d\theta}$ as well as via the limits of 
integration.

Fitting this function to the measured ${\theta}$ projection of the crystal map 
gives the position of the crystal vertices as seen by the chambers. 
A simple geometric transformation allows us to calculate the longitudinal
($z$) displacement of the chambers with respect to the CsI sphere. 
The resulting fits for the four different types of crystals are shown 
as examples in Fig.~\ref{fig:pos}.

The final result for a set of runs under study yields
\begin{equation}
      \Delta z =( 0.063 \pm 0.040 )\,{\rm mm},
\end{equation}
thus assuring us of the very precise alignment of the MWPC's and the CsI Calorimeter.

\bigskip
\section{PIBETA detector active target}\label{sec:tgt}
\medskip

Our regular active target is a cylindrical plastic scintillation counter 
with a length of 50$\,$mm and a diameter of 40$\,$mm. The counter is segmented into 
9 elements, as shown in a cross sectional drawing of the detector's central
region in Fig.~\ref{fig:central}. A central target tube
with a 3.5$\,$mm radius is surrounded with four identical quadrant segments 
that compose an inner ring with 15$\,$mm outer radius. The second, outer target 
ring is made out of four identically sized tubular segments that are rotated by
45$^\circ$ with respect to the inner ring.
These 9 pieces are wrapped individually in aluminized Mylar foil 
making them optically insulated from each other. The parts are pressed together
with a rubber band and the whole assembly is wrapped with
a black plastic tape. Each target element is viewed by a miniature (8$\,$mm 
diameter photo-cathode) Hamamatsu R7400U photomultiplier tube 
via a 60$\,$mm long, tapered acrylic light-guide and is therefore acting as 
an independent counter. A photograph of the partially assembled target detector
is presented in Fig.~\ref{fig:photo}. 

The analog signal from each of the target segments is divided by custom-made passive 
splitters into two branches. One side is discriminated in the time-over-threshold 
discriminator (CAMAC PS7106); its outputs are digitized in a Time-to-Digital 
Converter (FASTBUS LCR1877 TDC), as well as counted in a scaler unit (CAMAC 
PS7132H) that is read out every 10$\,$s. The second branch is connected to 
an Analog-to-Digital Converter (FASTBUS LCR1882F ADC), which is gated with 
a 100$\,$ns event trigger gate. 

High voltages of the target segments' PMTs are adjusted using a 4.1$\,$MeV $\pi^+\to 
\mu^+ \nu_\mu$ stopping muon line as a reference until the 9 detector gains are matched
when viewed on a digital oscilloscope. The final gain adjustment is done off-line
using the ADC spectrum of stopping pions (see the next Section).

The five central target detectors by design receive practically the entire stopped 
$\pi^+$ beam, each counting at a comparable rate.  The counting asymmetry 
(left--right or up--down) causes a shift of the centroid, that leads to a change 
in the integrated detector acceptance ratio.  This is an easily measurable asymmetry
even after the Poisson corrections for unresolved double hits and accidental 
coincidences with Michel decays.

\section{Beam profile from active target ADC spectra}\label{sec:adc}
\medskip

The analysis presented in this Section is based on the PIBETA data 
collected in 50 successive production runs, corresponding to about 2 days of 
uninterrupted data taking at the nominal $\pi^+$ stopping rate of
$\simeq 5\cdot 10^5\,$s$^{-1}$.

Two sets of ADC histograms for each of our two beam triggers are defined for 
9 individual target segments. These histograms are filled off-line with 
an additional requirement that no particle showers are detected in the CsI 
calorimeter, thus suppressing hadronic interaction events and delayed
weak $\pi^+$ or $\mu^+$ decays with a final state positron, either of
which could affect the target energy depositions.

Representative energy spectra for one target piece (No. 1) are
shown in Fig.~\ref{fig:pi_e_tgt}. The events in which the stopping pion 
(or thru-going positron) deposits its full energy in this target element are 
clearly distinguished in the high-energy peak. The energy spectra of all 
9 target elements are 
empirically well represented by Gaussian distributions with 
square-root function tails:
\begin{equation}
f(E)=p_1 exp\Bigg[ {-{1\over 2} \Bigg( {{E-p_2}\over {p_3}} \Bigg) ^2} \Bigg]+
\theta(p_4-p_5 E) \sqrt{p_4-p_5 E},
\label{eq:2d}
\end{equation}
where $p_i$ are the 5 free parameters of the fit.
The events with the stopping particle sharing energy between two or more target 
segments due to scattering and/or beam spreading fall into a low energy tail.
The low energy tail is also populated by positrons from weak pion and muon decays
which miss the calorimeter. The peak-to-tail amplitude ratio depends on the pion
stopping rate, the ADC gate width, as well as on the divergence of the beam at 
the front face of the target. The comparison of the data with the target energy 
depositions calculated with GEANT3 Monte Carlo~\cite{Bru94} suggests that the beam 
divergence $p_r/p_z$ is small, $\simeq\,$\,1\% (see also Sec.~\ref{sec:dal}).

We have extracted the total number of pions and positrons stopping in each 
target segment by fitting their high-end energy spectra with Gaussian functions,
integrating the fitted peaks and normalizing the integrals to the Monte Carlo
predictions. The results are summarized in Fig.~\ref{fig:tgt9}.
The top panel shows the target counting rates for the ``$\pi$-in-beam'' trigger, while 
the bottom diagram corresponds to the ``$e$-in-beam'' events. The standard deviations 
shown are calculated from the uncertainties of the fitted Gaussian parameters and 
reflect both the event statistics and the quality of the fits.
The central segment and the inner ring $\pi^+$ counting rates are determined with 
an accuracy better than 1\%. The outer ring segments count the tails of the beam
distribution with a 3--5\,\% uncertainty.

We have examined the resulting target counting rates by simulating the target's response 
to the variety of lateral stopping beam distributions with the help of a simple Monte Carlo 
(MC) integration program. Good agreement between the Monte Carlo simulation
and the measured rates is achieved by approximating the 3-dimensional beam profile with 
three separate Gaussian distributions in the three orthogonal variables $x$ (beam left),
$y$ (up), and $z$ (beam downstream):
\begin{eqnarray}
f(x,y,z)= N_0\Bigg\{ && \theta(x_0-x)\cdot exp\Bigg[ {-{1\over 2} \Bigg( {{x-x_0}\over 
{\sigma_{xL}}} \Bigg) ^2} \Bigg] +\nonumber \\ 
                     &&   \theta(x-x_0)\cdot exp\Bigg[ {-{1\over 2} \Bigg( {{x-x_0}\over 
{\sigma_{xR}}} \Bigg) ^2} \Bigg]
         \Bigg\} \\ 
            &&  exp\Bigg[ {-{1\over 2} \Bigg( {{y-y_0}\over {\sigma_y}} \Bigg) ^2}\Bigg]
           \cdot exp\Bigg[ {-{1\over 2} \Bigg( {{z-z_0}\over {\sigma_z}} \Bigg) ^2}\Bigg]
         \nonumber ,  
\label{eq:eq}
\end{eqnarray}
where the function $\theta(x)=0$ for $x>0$ and $\theta(x)=1$ for $x>0$, $\sigma_{xL}$ and $\sigma_{xR}$ 
are standard deviations of the horizontal $x$ beam profile beam left ($\rm B_L$) and beam right 
($\rm B_R$), $\sigma_{y}$ and $\sigma_{z}$ are beam widths in the vertical ($y$) and axial ($z$)
projections, and the beam spot is centered at the point $(x_0,y_0,z_0)$. 
The asymmetry in the horizontal ($x$) beam profile is caused by a $4\,$mm thick carbon degrader 
inserted in the middle of the $\pi$E1 beam-line. The momentum-analyzed pions and positrons have 
different energy losses in the carbon absorber, and are therefore spatially separated 
in the horizontal plane. The passive lead collimator in front of the PIBETA detector
then blocks out most $e^+$'s and $\mu^+$'s, minimizing the non-pionic beam contamination.

All other attempted descriptions of the relative counting rates (e.g., a Lorentzian, 
piecewise parabolic functions, a box spectrum with or without exponential tails, etc.) 
resulted in far worse fits of the data.

The five parameters of the $f(x,y)$ distribution, namely $\sigma_{xL}$, $\sigma_{xR}$,
$\sigma_y$, $x_0$, and $y_0$ are calculated using the MINUIT 
minimization code~\cite{Jam89} by requiring the best agreement between the distribution
(2) and the data in Fig.~\ref{fig:tgt9} (minimum $\chi^2$). 
The SEEK and SIMPLEX search methods are used to suppress the dependence 
of the convergence on the uncertainty of the Monte Carlo integration. 
SEEK minimizes the function using the Metropolis algorithm 
by selecting random values of the variable parameters, chosen uniformly 
over a hypercube centered at the best current value. SIMPLEX is 
a genuine multidimensional function minimization routine robust even 
to gross fluctuations in the function value. Both algorithms converged 
to the same solution. 

The best parameter values obtained in the fits are summarized in the second column
of Table~\ref{tab:sum}. The corresponding two-dimensional distribution is shown in 
Fig.~\ref{fig:extr_2d}, where it is superimposed on the sub-division of our target. 
The agreement between the measured counting rates and 
the simulated rates is within one standard deviation, with 
the $\chi^2$ per degree of freedom of 0.9.

\bigskip
\section{{\boldmath $\pi^+$} beam profile from back-tracking tomography}\label{sec:tomo}
\medskip
\subsection{Longitudinal $\pi^+$ stopping distribution}\label{sec:long}
\medskip
The thicknesses of the beam-defining detectors, namely of the forward
beam counter, the active degrader and the active target, were chosen
to make the 40.6$\,$MeV incident $\pi^+$ beam particles stop close to
the center of the target. Our GEANT3 Monte Carlo 
simulation of the longitudinal vertex distribution of decaying
$\pi^+$'s is presented in Fig.~\ref{fig:pistop_z}.
The input to the Monte Carlo is the $\pi^+$ momentum spectrum in 
Fig.~\ref{fig:beam_p}. A histogram of Monte Carlo $z$ coordinates of 
$\pi^+$ decay vertices is a Gaussian function with a width of 
$\sigma_z=1.69\pm0.01\,$mm and a flat upstream tail integrating
to $0.86\pm 0.05$\,\% events. The $z$ position spread originates 
mainly from the energy straggling of stopping pions: 
the momentum spread of the incident beam contributes just 
$0.2\,$mm (or 12\%) to the over-all axial distribution spread.  
The upstream tail represents $\pi^+$ decay-in-flight events
and strong interactions with detector material.
\medskip
\subsection{MWPC tracking and positional resolution}\label{sec:mwpc}
\medskip
The main decay channel of $\pi^+$'s stopped in the target is the $\pi^+\to\mu^+\to e^+$ 
decay chain:
\begin{equation}
\pi^+\to \mu^+\nu_\mu,
\label{eq:pumu}
\end{equation}
followed by
\begin{equation}
\mu^+\to e^+\nu_e\bar{\nu}_\mu.
\label{eq:mich}
\end{equation}
The first reaction, namely the main two-particle pion decay, results in 
a monoenergetic $\mu^+$ with a kinetic energy of 4.1$\,$MeV that stops within 
the target detector because of its short 1.34$\,$mm range. 
The muons subsequently decay, by producing so-called Michel positrons with 
a typical continuous $\beta$-decay energy spectrum characterized by its 52.5$\,$MeV 
($=m_\mu/2$) end-point. A GEANT3 simulation predicts that only $\sim$\,0.48\,\% of 
the $\mu^+$'s escape the target before coming to a complete stop.

The suppressed $\pi_{e2}$ decay mode
\begin{equation}
\pi^+\to e^+\nu_e
\label{eq:p2e}
\end{equation}
proceeds with the branching ratio $B_{\pi_{e2}}=1.24\cdot 10^{-4}$~\cite{PDG}
and results in a monoenergetic 69.8$\,$MeV positron that deposits
an average of 4.5$\,$MeV in the target. All average deposited energies, particle ranges
and losses calculated in simulation assume $\sigma_{x,y}\simeq 8\,$mm nominal beam spot size
centered on the target.  

Michel positrons [Eq.~(\ref{eq:mich})] and $\pi\to e\nu$ positrons  [Eq.~(\ref{eq:p2e})] 
are identified in the segmented plastic veto hodoscope PV via their characteristic energy
depositions and tracked in the pair of 
cylindrical MWPC's~\cite{Kar98} that surround the target region. 
Total positron energies are measured in the CsI calorimeter where they
produce fully contained showers. The particles' energy spectra in the CsI 
calorimeter are used to discriminate between $\pi^+$ and $\mu^+$ decay processes.
The $\pi_{e2}$ positron vertices in the target therefore coincide 
with the $\pi^+$ stopping points. The Monte Carlo root-mean-square (rms) of 
the Michel vertices turns out to be merely 0.04\,mm larger than the $\pi_{e2}$ rms.

The directional resolution of the MWPC's has to be known before one
can use back-tracking of the charged particles to find the real distribution
of vertices in the target. Cosmic muon events are ideally
suited for the resolution study of the MWPC's. Clean samples of
cosmic muon events are collected during weekly beam-off 
cyclotron maintenance periods. The cosmic event trigger requires
the coincidence of the cosmic muon veto counter and a CsI calorimeter signal 
above the ``low discriminator threshold'' ($\sim$\,5\,MeV). 

In the off-line data analysis two additional stringent software cuts 
are imposed: (i) CsI calorimeter energy deposition $>$\,200\,MeV, and
(ii) exactly two reconstructed hits in each MWPC. These constraints 
effectively remove any extraneous non-cosmic muon background.
We reconstruct the cosmic muon track from a pair of hits in one MWPC 
and calculate the intersections of that track with the other chamber.
Histogrammed differences between the calculated $x_c$, $y_c$, $z_c$
and measured intersection coordinates $x_e$, $y_e$, and $z_e$ represent 
the positional resolutions of the chamber. The typical azimuthal angle 
resolution $\Delta \phi$ and axial coordinate resolution $\Delta z$ for 
the MWPC$_1$ chamber are shown as an illustration in two panels of 
Fig.~\ref{fig:mwpc_r}. The $\Delta \phi$ rms is 0.74$^\circ$ and
$\Delta z$ rms is 0.97\,mm.

\medskip
\subsection{Back-tracking tomography: reconstruction algorithm}\label{sec:alg} 
\medskip
The tomography algorithm relies on the back-tracking data obtained with the MWPC's. 
In the off-line analyzer computer program we define an $80\times 80\times 80\,$mm$^3$ volume 
containing the cylindrical beam stopping target divided into elementary cubic 
cells of fixed size. We use initially 80$\times$80$\times$80 elementary cells, 
each with a volume of 1\,mm$^3$. For each reconstructed 
$\pi_{e2}$ event a positron path length inside each elementary cell that 
is intersected by a MWPC track is found first. The analyzer
code keeps a cumulative sum of tracks' path lengths for each individual cell. 
The high statistics data set ensures that the cubic cell containing 
the most $\pi_{e2}$ vertices accumulates the largest path length sum.
The distribution of path lengths therefore reflects the distribution 
of the decaying particles' vertices originating from the different volume 
elements~\cite{Gor74}.

The above-described algorithm determines accurately the center position 
$(x_0,y_0,z_0)$ of the $x_{\pi^+}$, 
$y_{\pi^+}$, and $z_{\pi^+}$ distributions. While the shape of our $\pi^+$ 
stopping distribution can be approximated by the product of Gaussian functions
(Sec.~\ref{sec:adc}), the real lateral distribution is more complicated, 
exhibiting outlying tails.

The Monte Carlo (MC) method is then used to find the correspondence between the input
track vertex distribution in the target and the output path length distribution
in the elementary cells. In the first stage one has to optimize the cell size.

We generate the MC tracks from an asymmetric Gaussian input distribution (Eq.~\ref{eq:eq})
uniformly into the solid angle covered by the MWPC geometry. The points at which 
the tracks intersect the cylindrical chambers are smeared next by the measured 
MWPC positional resolutions. We simulate the experimentally measured particle direction 
by finding the track through smeared intersection points associated with the two MWPC's. 
Processing the large statistics of the MC tracks obtained in this way with
the tomography algorithm described above we simulate the $\Sigma x_T$, $\Sigma y_T$ and 
$\Sigma z_T$ projections of the path length sums. The Gaussian fits to the simulated
path length histograms are thus connected to the input distribution rms's.
The dependence of the reconstructed width of $\Sigma y_T$ on the unit cell size is
shown in Fig.~\ref{fig:tomosys}. The input rms value of the vertical beam profile 
was 10\,mm. From this study we deduce that for an optimal cell size of 0.3\,mm 
a small correction of $\Delta \sigma_y=0.008\pm 0.005\,$mm has to be applied to 
the extracted rms of the vertical ($y$) vertex distribution.

The horizontal/lateral ($x$) and axial ($z$) rms values of the vertex profiles
are treated analogously and the small correction factors ($< 0.1\,$mm)
corresponding to the chosen cell size are obtained. The two-dimensional ``contour'' 
plot in Fig.~\ref{fig:lookup} shows the correspondence between the input vertex 
distribution in the horizontal coordinate $x$ and the simulated path length histogram. 

The MC tomography reconstruction is compared with the experimental data in
three panels of Fig.~\ref{fig:tomo}. The agreement between the
data points and the predicted projections is very good: 
by scaling the width of the input $f(x,y)$ vertex distribution and studying
the changes in the resulting chi-squares we estimate the uncertainty in the extracted 
$\sigma_{x_{\pi^+}}$ and $\sigma_{y_{\pi^+}}$ to be $\sim$\,0.05\,mm.
These results are summarized in the third column of Table~\ref{tab:sum}.
The details of the analysis are given in Ref.~\cite{Li04}.
\bigskip

\section{Pion beam profile: Dalitz event {\boldmath $e^+e^-$} vertices}\label{sec:dal}
\medskip
The selection criteria for $\pi^0$ Dalitz decay events, $\pi^0\to e^+e^-\gamma$,
were:
\begin{itemize}
\item[(1)] no in-time hits in either active collimators or in cosmic 
muon vetoes;
\item[(2)] no prompt $\pi^+$ in either the forward beam counter or in 
the active degrader;   
\item[(3)] total calorimeter energy deposition $10\,{\rm MeV}<E_C<200\,$MeV;
\item[(4)] at least two  minimum ionizing (MI) particle tracks pointing back
to the target volume, identified as $e^\pm$'s via energy depositions 
in the plastic veto and CsI calorimeter;
\item[(5)] at least one neutral shower with $E_\gamma>20\,$MeV in the CsI 
calorimeter within $\pm 5\,$ns of two MI tracks;
\item[(6)] if two or more candidate $e^\pm$ and/or $\gamma$ tracks
are found in a single event, the $(\gamma,e^+,e^-)$ combination with the closest timing 
is selected.
\end{itemize}   
The candidate events satisfying the conditions listed above were
written into a data summary tape in the off-line data replay.

The $e^+$ and $e^-$ tracks were back-projected to the target region
and a pair of points, one on each track, which are closest to
each other was found. The mid-point coordinates on the line connecting 
the closest two track points are defined as an event vertex.

Fig.~\ref{fig:dal} shows the projection of Dalitz event vertices
onto the horizontal $(x)$, vertical $(y)$, and beam ($z$) axes. The extracted 
peak widths, defined as a beam right and beam left Gaussian fits to the vertex
distributions, are entered in Table~\ref{tab:sum}, fourth column.

The widths of the beam stopping distribution as well as its position
are in excellent agreement with the tomography solution discussed in 
Sec.~\ref{sec:alg} above.

\bigskip

\section{{\boldmath $\pi^+$} beam profile and {\boldmath $\pi^0\to \gamma\gamma$}
opening angle}\label{sec:ang}
\medskip
We have developed a complete GEANT3 Monte Carlo description of 
the PIBETA detector that includes all major sensitive as well as passive 
detector components. The user code is written in a modular 
form in standard FORTRAN 77 and organized into over 300 subroutines 
and data files~\cite{Frl97}.

The lateral distribution of the $\pi\beta$ event vertices in the $x$--$y$ 
plane in our simulation is derived from the tomographic beam spot reconstruction
of Fig.~\ref{fig:tomo}.
In the direction of the beam axis the Gaussian $\pi^+$ stopping 
spectrum is generated with $\sigma_z=1.7\,$mm, as shown in 
Fig.~\ref{fig:pistop_z}.

The MC $\pi\beta$ event generator takes into account the
phase space probability, the square of the matrix element
and the radiative corrections~\cite{Kal60,Gin66}. 
The GEANT3 $\pi\beta$ detector trigger relies on the response 
of the CsI calorimeter; it requires two coincident neutral
showers above the high energy discriminator threshold (HT) of 51.5$\,$MeV
that reconstruct into a $\pi^0$ originating from the $\pi\beta$ decay
at rest.

The photon directions calculated by a logarithmic weighting
method of shower localization~\cite{Awe92}. The method is
motivated by the exponential fall-off of the shower's transverse
energy deposition~\cite{Bug86}. The angular direction $(\theta_c,\phi_c)$ 
of a particle initiating an electromagnetic shower in the PIBETA calorimeter 
is found as an energy-weighted mean:
\begin{equation}
\theta_c = { {\sum\limits_{i=0}^{N} \omega_i(E_i)\theta_i}\over
{\sum\limits_{i=0}^{N} \omega_i(E_i)} },
\end{equation}
where $\theta_i$ is the polar angle of
an individual CsI module, $\omega_i$ is a weighting function, and $N$ is 
the number of nearest neighbors of the crystal with maximum energy deposition 
$E_0$. The sum is over the central crystal ``0'' and all of its nearest
neighbors and thus contains 6 to 8 terms, 
depending on the shape of the centrally hit module. The formula for the azimuthal
angle $\phi_c$ is identical to that for the polar angle with the replacement
$\theta_i \to \phi_i$.

The weighting function 
$\omega_i$ is defined by:
\begin{equation}
\omega_i=\mbox{max}\big[ 0, a_0+\ln(E_i)-\ln(E_{tot})\big],
\end{equation}
where the best value for the constant $a_0$ is determined empirically 
to be 5.5, and $E_{tot}$ is the total energy deposited in 
$N+1$ modules that define the shower cluster.

The shape and absolute integral of the pion beta decay $\gamma$-$\gamma$
opening angle histogram is very sensitive to the distribution of vertices 
of the initial decaying $\pi^0$'s. The acceptance $A_{\pi\beta}^{\rm HT}$
for the high discriminator threshold HT calculated in a GEANT3 Monte Carlo 
simulation is
\begin{eqnarray}  
A_{\pi\beta}^{\rm HT} &=&
(1-0.4928\cdot 10^{-3}\cdot r_0 -0.7365\cdot 10^{-3}\cdot r_0^2)\times
\nonumber \\
 & & (0.5977-0.2350\cdot 10^{-2}\cdot s_r -0.5500\cdot 10^{-3}\cdot s_r^2),
\label{eq:11}
\end{eqnarray}  

where $r_0=\sqrt{x^2+y^2}$ is a lateral offset of the beam center, and $s_r$ is 
a lateral width (a standard deviation) of a Gaussian-shaped beam spot,
where both variables are expressed in millimeters. This expression
represents our main result, parameterizing the acceptance of
the PIBETA detector.

Fig.~\ref{fig:th_gg} shows
$\gamma$-$\gamma$ opening-angle histograms calculated using a GEANT3 
code for three different beam spot sizes (top panel) and varying axial positions 
of the beam center (middle panel). The figure indicates that it pays to keep 
the beam spot root-mean square spread $s_r$ as small as possible, because 
the detector acceptance for a tight event vertex distribution is less sensitive 
to its shape and position. 
In particular, the integrated $\pi\beta$ event acceptance is relatively insensitive to
the longitudinal position $z_0$ of the stopping point, with $A_{\pi\beta}^{\rm HT}$
changing $<$0.07\,\% for $z_0=\pm 10$\,mm, and is therefore neglected in Eq.~\ref{eq:11}.

On the other hand, the resulting $\pi\beta$ acceptance is affected considerably by 
the spreading and lateral offsets of the beam profiles. Using the 
parameterization of Eq.~(\ref{eq:11}) and propagating the uncertainties of the beam spot 
variables $r_0$ and $s_r$ we obtain the results in Table~\ref{tab:err}.
At our nominal stopped $\pi^+$ beam width of $\sigma_{x,y}\simeq 8$\,mm, 
the change in the rms value or lateral position of the beam spot by 
$\simeq$\,1\,mm affects the $\pi\beta$ acceptance by $\simeq$\,1.5\,\%. Therefore, 
the experimental knowledge of the beam geometry at the sub-millimeter scale 
is required to accomplish the goals of the PIBETA experiment. This paper demonstrates 
that the required accuracy has been achieved. If absolute normalization is done 
with respect to the $\pi_{e2}$ acceptance, the constraint on the beam 
geometry uncertainty is not changed significantly because in our experimental configuration
$A_{\pi_{e2}}$ is almost independent of the small variations from 
the nominal beam model. The geometrical acceptance $A_{\pi_{e2}}$ varies
less than $0.5$\,\% for $r_0$, $z_0$, and $s_r$ in 0--10\,mm range.

In the bottom panel of Fig.~\ref{fig:th_gg} we present the MC 
$\theta_{\gamma\gamma}$ calculation for our average beam distribution. 
The measured $\theta_{\gamma\gamma}$ spectrum, based on an off-line replay 
of 59,011 $\pi\beta$ events is represented by full markers. For data 
taken over such a long time period we must take into account the temporal 
stability of the beam distribution. The back-tracing tomography
reveals that our beam spot position and its size were remarkably stable. 
The reconstructed lateral ($r_T$) and longitudinal ($z_T$) centroids of 
the stopping distribution vary with
rms$_r\simeq$\,0.04\,mm and rms$_z\simeq$\,0.2\,mm, respectively,
as shown in Fig.~\ref{fig:beam_stab}. The oscillations of the 
beam spot location and size are included in our GEANT3 event
generators. After that adjustment, $\pi\beta$ data points and the MC simulation of 
$\theta_{\gamma\gamma}$ agree with each other within the statistical 
uncertainties.

\bigskip
\section{Conclusion}\label{sec:conc}
\medskip
Precise knowledge of the PIBETA detector acceptance for 
the decay products of the stopped pion beam is required
for the evaluation of the absolute branching ratios of pion 
and muon decay processes. The geometrical acceptances of a compact 
detector like ours, defined with the specific cuts on the energies 
and angles of measured particles, depend strongly on the distribution 
of vertices of decaying particles. In this paper we show
that we can measure position and the volume distribution
of the stopped pion beam with sub-millimeter precision.
The uncertainty in the 3-dimensional beam shape or position of the order 
of $\simeq$\,0.1\,mm then translates into $\simeq$\,0.12\,\% systematic 
uncertainty of the acceptance calculated with the GEANT3 Monte 
Carlo simulation. The consistency of our four different methods
of extracting the pion beam stopping distribution is summarized
in Table~\ref{tab:sum}. 

\bigskip
\section{Acknowledgements}
\medskip
We thank J.~Koglin who developed the TURTLE and TRANSPORT MC simulations
of our beam tunes. The mechanical support
of the {\sc PIBETA} detector was designed and built under supervision of
H.~Obermeier. Z.~Hochman provided valuable technical assistance
throughout the detector assembly and commissioning phases. The support from
the Hallendienst and many other PSI staff members is gratefully acknowledged.

This work is supported and made possible by grants from the US National
Science Foundation and the Paul Scherrer Institute.


\clearpage

\vspace*{\stretch{1}}
\begin{figure}[!tpb] 
\caption{The momentum spectrum of the beam $\pi^+$'s entering the active 
degrader/target counters calculated by the TURTLE program~\protect{\cite{Bro74}}.}
\label{fig:beam_p}
\end{figure}

\begin{figure}[!tpb] 
\caption{A ``$\pi$-in-beam'' trigger is defined as a four-fold coincidence
between a forward beam counter (Ch. 1), an active degrader (Ch. 2), an active
target (Ch. 3), and a 19.75$\,$ns rf cyclotron signal (Ch. 4).}
\label{fig:pistop}
\end{figure}

\begin{figure}[!tpb] 
\caption{Determination of the offsets between coordinate systems of
the MWPC's and CsI calorimeter. The MWPC's track intersections with
the calorimeter sphere for which four representative crystals also have
the maximum energy depositions.}
\label{fig:pos}
\end{figure}

\begin{figure}[!tpb] 
\caption{Cross section of the PIBETA detector central region (drawn in GEANT3).
Shown are the segmented active target AT, a pair of cylindrical MWPC's,
and the plastic veto hodoscope array PV.}
\label{fig:central}
\end{figure}

\begin{figure}[!tpb] 
\caption{A photograph of the segmented PIBETA active target. Tapered acrylic
light-guides are glued to the mutually optically isolated 
target segments.}
\label{fig:photo}
\end{figure}

\begin{figure}[!tpb] 
\caption{Representative energy spectra of beam pions 
($\pi$-in-beam trigger, top panel) and beam positrons ($e$-in-beam 
trigger, bottom panel) stopping in the target segment number 1. 
Monte Carlo GEANT3 predictions are represented by full line spectra.}
\label{fig:pi_e_tgt}
\end{figure}

\begin{figure}[!tpb] 
\caption{Integrated counting rates in 9 target segments for the
$\pi$-in-beam and $e$-in-beam triggers, respectively. Orientation
of the coordinate system and target segment labels (0--8) are also
indicated.}
\label{fig:tgt9}
\end{figure}

\begin{figure}[!tpb] 
\caption{The 2-dimensional shape of the $\pi^+$ beam
superimposed on an outline of the segmented target. The reconstruction is
based on the target ADC spectra collected during years 1999--2001.}
\label{fig:extr_2d}
\end{figure}



\begin{figure}[!tpb] 
\caption{The $z$-component of the $\pi^+$ stopping distribution in 
the active target calculated in the GEANT3 simulation using 
the momentum distribution in Fig.~\ref{fig:beam_p}.}
\label{fig:pistop_z}
\end{figure}

\begin{figure}[!tpb] 
\caption{The MWPC$_1$ directional resolutions in the azimuthal angle $\phi$ (top panel), 
and axial coordinate $z$ (bottom panel) extracted from a sample of cosmic muon events.
The data points for $\phi$ and $z$ coordinates are fitted with a sum of two Gaussians, 
the broader one accounting for the cosmic muon scattering in the apparatus.}
\label{fig:mwpc_r}
\end{figure}

\begin{figure}[!tpb] 
\caption{The dependence of reconstructed vertical rms widths $\sigma_y$ of the beam 
stopping distribution on the size of the elementary cell used in the calculation.}
\label{fig:tomosys}
\end{figure}

\begin{figure}[!tpb] 
\caption{A lookup table for determining $\sigma_{xL/R}$ ($L$ left, $R$ right) is 
calculated starting from the synthetic source distributions in the Monte Carlo. 
The fitted ``contour'' lines correspond to the values of equal $\sigma$'s.}
\label{fig:lookup}
\end{figure}

\vspace*{\stretch{2}}
\clearpage
\vspace*{\stretch{1}}

\begin{figure}[!tpb] 
\caption{The distribution of path length sums for $x$ (top), $y$ (middle) and $z$
(bottom) coordinate projections. The experimental points for back-tracked 
$\pi^+\to e^+\nu_e$ positrons are compared with the Monte Carlo simulation that uses the beam 
stopping profile deduced from back-tracking tomography.}
\label{fig:tomo}
\end{figure}

\begin{figure}[!tpb] 
\caption{The horizontal ($x$, the top panel) and vertical ($y$, the middle
panel) beam profiles deduced from the Dalitz event $\pi^0\to \gamma e^+e^-$
vertices. The axial $z$ coordinate cut around the target center, $\Delta 
z=\pm 10\,$mm is imposed to select the $\pi^+$ stopping distribution
(the bottom panel). The agreement with the tomography solutions is
very good.}
\label{fig:dal}
\end{figure}

\vspace*{\stretch{1}}
\begin{figure}[!tpb] 
\caption{The $\gamma$-$\gamma$ opening angle following the $\pi^0$
decay from the $\pi^+\to \pi^0 e^+\nu$ process as reconstructed 
in the PIBETA segmented calorimeter. The GEANT3 prediction for 
the stopping $\pi^+$ beam profiles with three different transverse 
widths (top panel) and three different longitudinal stopping points
(middle) is shown. The bottom panel shows a comparison of the partial
$\pi\beta$ events data set with the Monte Carlo prediction. The plotted
uncertainties are statistical only.}
\label{fig:th_gg}
\end{figure}

\begin{figure}[!tpb] 
\caption{The spatial stability of the $\pi^+$ beam spot center 
during one three month data taking period. The lateral beam center 
location (top) and the axial beam stopping point (bottom) are 
reconstructed via back-tracking tomography. The reconstruction 
uncertainty for a single run is $<$0.02\,mm.}
\label{fig:beam_stab}
\end{figure}


\vspace*{\stretch{2}}
\clearpage

\bigskip
\begin{table}[ph]
\caption{The rms widths and centering of the $\pi^+$ beam spot in the PIBETA
active stopping target deduced by four different methods: (i) 
using the counting rate matrix in the segmented target, (ii) via
back-tracking of $\pi^+\to \mu^+\to e^+$ positrons with two cylindrical
MWPC's, (iii) from Dalitz decay $e^+e^-$ pair vertices, and (iv) from fits 
to the opening angle distribution of two $\pi^0\to \gamma\gamma$ photons. 
BR/BR stands for beam right/left, BU/BD labels beam up/down directions.}
\bigskip
\label{tab:sum}
\begin{tabular}{lcccc}
\hline
\multicolumn{1}{c}{Beam profile}
&\multicolumn{1}{c}{Segmented} 
&\multicolumn{1}{c}{MWPC's}
 &\multicolumn{1}{c}{Dalitz $e^+e^-$}
&\multicolumn{1}{c}{$\theta_{\pi^0\to \gamma\gamma}$}\\
\multicolumn{1}{c}{parameters} 
&\multicolumn{1}{c}{target rates}
&\multicolumn{1}{c}{back-tracking}
&\multicolumn{1}{c}{vertices} 
&\multicolumn{1}{c}{opening angle}\\
\hline\hline
$\sigma_x^R$ (mm, BR horizontal) & $7.0\pm 0.3$ & $7.04\pm 0.04$ & $6.67\pm 0.37$ & $7.0\pm 0.5$ \\
$\sigma_x^L$ (mm, BL horizontal) & $8.8\pm 0.3$ & $9.50\pm 0.07$ & $8.46\pm 0.75$ & $9.0\pm 0.5$ \\
$\sigma_y^U$ (mm, BU vertical)   & $7.0\pm 0.3$ & $7.44\pm 0.06$ & $7.25\pm 0.49$ & $7.6\pm 0.5$ \\
$\sigma_y^D$ (mm, BD vertical)   & $7.0\pm 0.3$ & $7.46\pm 0.07$ & $7.40\pm 0.32$ & $7.6\pm 0.5$ \\
$\sigma_z$ (mm, axial)           & --           & $1.70\pm 0.03$ & $1.8\pm 0.3$   & $2.0\pm 0.3$ \\
$x_0$ (mm)                       & $-1.0\pm 0.3$& $-1.92\pm 0.05$ & $-0.95\pm 0.70$ & $-1.0\pm 0.3$\\
$y_0$ (mm)                       & $0.1\pm 0.3$ & $-0.12\pm 0.05$ & $-0.31\pm 0.60$ & $0.0\pm 0.3$\\
$z_0$ (mm)                       & --           & $0.30\pm 0.02$  & $0.43\pm 0.12$ & $0.30\pm 0.1$\\
\hline
\end{tabular}
\end{table}

\clearpage

\bigskip
\begin{table}[ph]
\caption{Summary of two main uncertainties connected with the size and position
of a $\pi^+$ beam spot in determining the detector acceptance $A_{\pi\beta}$. Relevant
geometrical variables are a lateral offset of the $\pi^+$ beam center $r_0$
with respect to the geometrical center of the target and a lateral rms width of the 
$\pi^+$ beam profile $s_r$. Beam spot parameters obtained from the back-tracking 
tomography are used.}
\bigskip
\label{tab:err}
\begin{tabular}{lcc}
\hline
\multicolumn{1}{c}{Source}
&\multicolumn{1}{c}{General}
&\multicolumn{1}{c}{Current}\\
\multicolumn{1}{c}{\ }
&\multicolumn{1}{c}{value (\%)}
&\multicolumn{1}{c}{value (\%)}\\
\hline\hline
$(A_{\pi\beta}^{\rm HT})^{-1}\cdot (\partial A_{\pi\beta}^{\rm HT}/\partial r_0)\cdot
   \sigma_{r_0}\vert _{r_0=1\,{\rm mm}}$ & $1.15\cdot \sigma_{r_0}$\,({\rm mm}) & 0.06 \\
$(A_{\pi\beta}^{\rm HT})^{-1}\cdot (\partial A_{\pi\beta}^{\rm HT}/\partial s_r)\cdot
   \sigma_{s_r}\vert _{s_r=8\,{\rm mm}}$ & $1.26\cdot \sigma_{s_r}$\,({\rm mm}) & 0.10 \\
\hline
\end{tabular}
\end{table}
\bigskip   


\clearpage

\vspace*{\stretch{1}}
\centerline{\psfig{figure=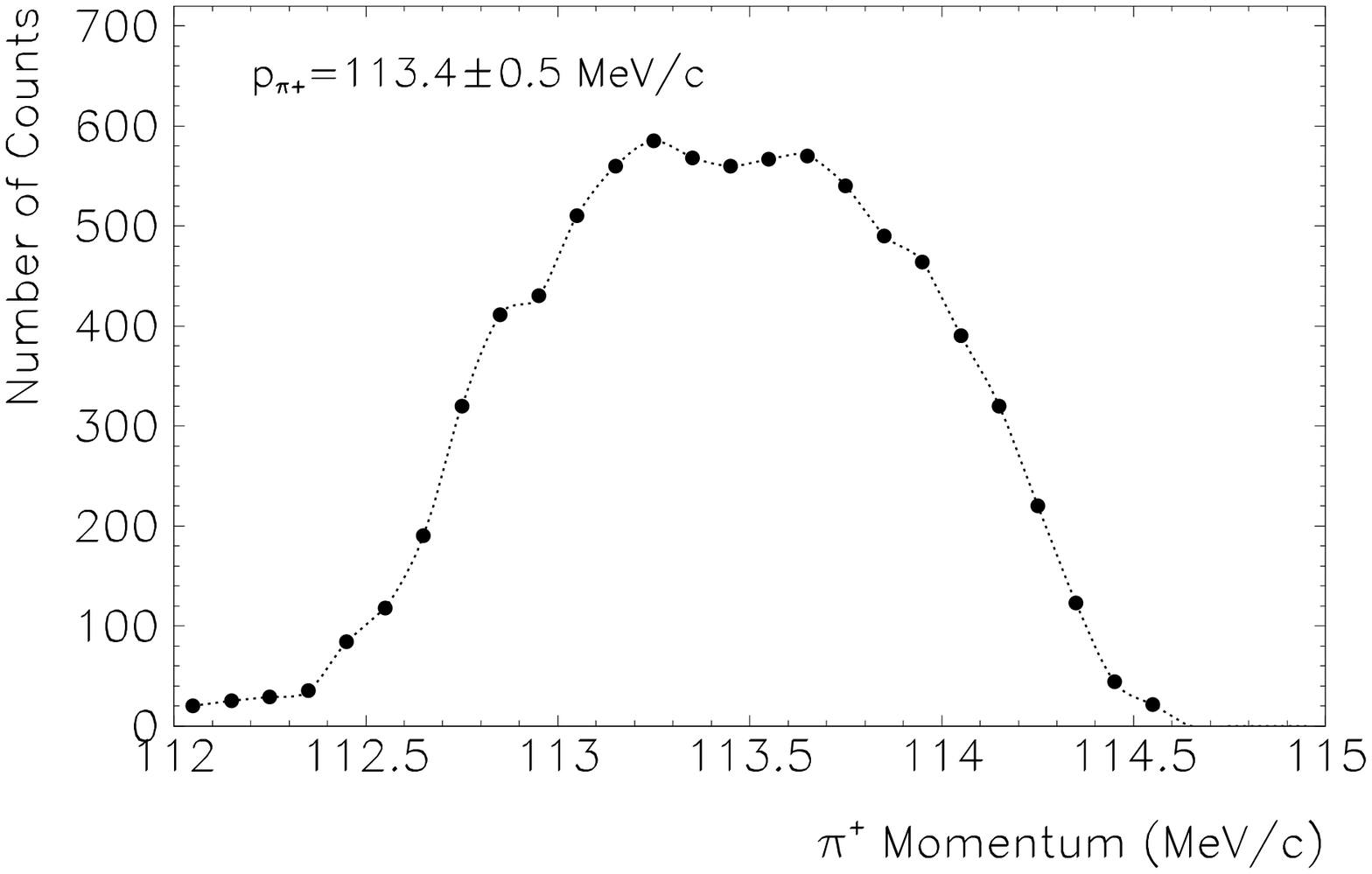,height=22cm}}
\vglue -9.5cm
\centerline{FIGURE~\ref{fig:beam_p}}
\vspace*{\stretch{2}}
\clearpage

\vspace*{\stretch{1}}
\centerline{\psfig{figure=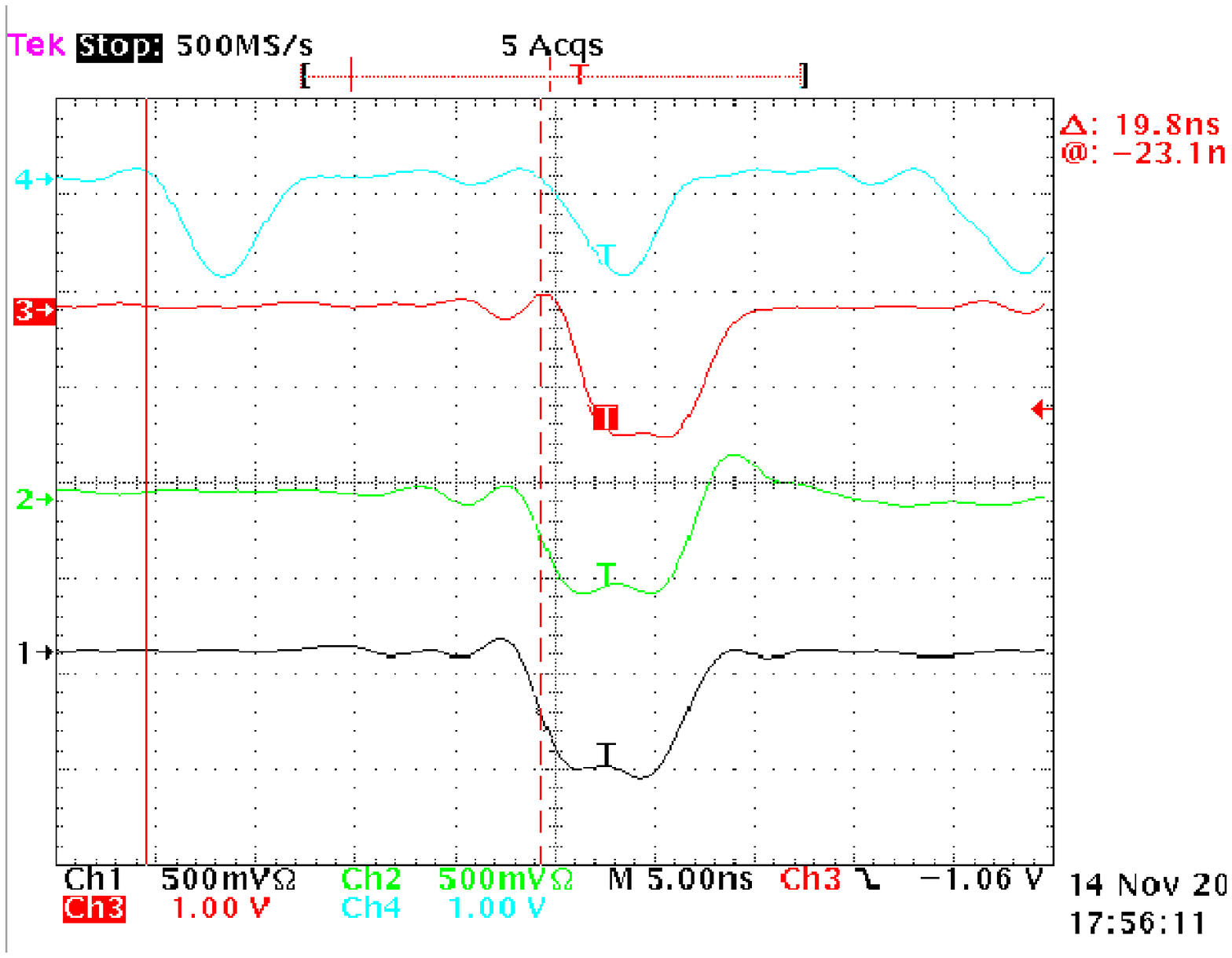,height=12cm}}
\bigskip\bigskip
\centerline{FIGURE~\ref{fig:pistop}}
\vspace*{\stretch{2}}
\clearpage

\vspace*{\stretch{1}}
\vglue -0.5cm
\hbox to \textwidth
 {\vbox{\hsize=3.1in
  \centerline{\psfig{figure=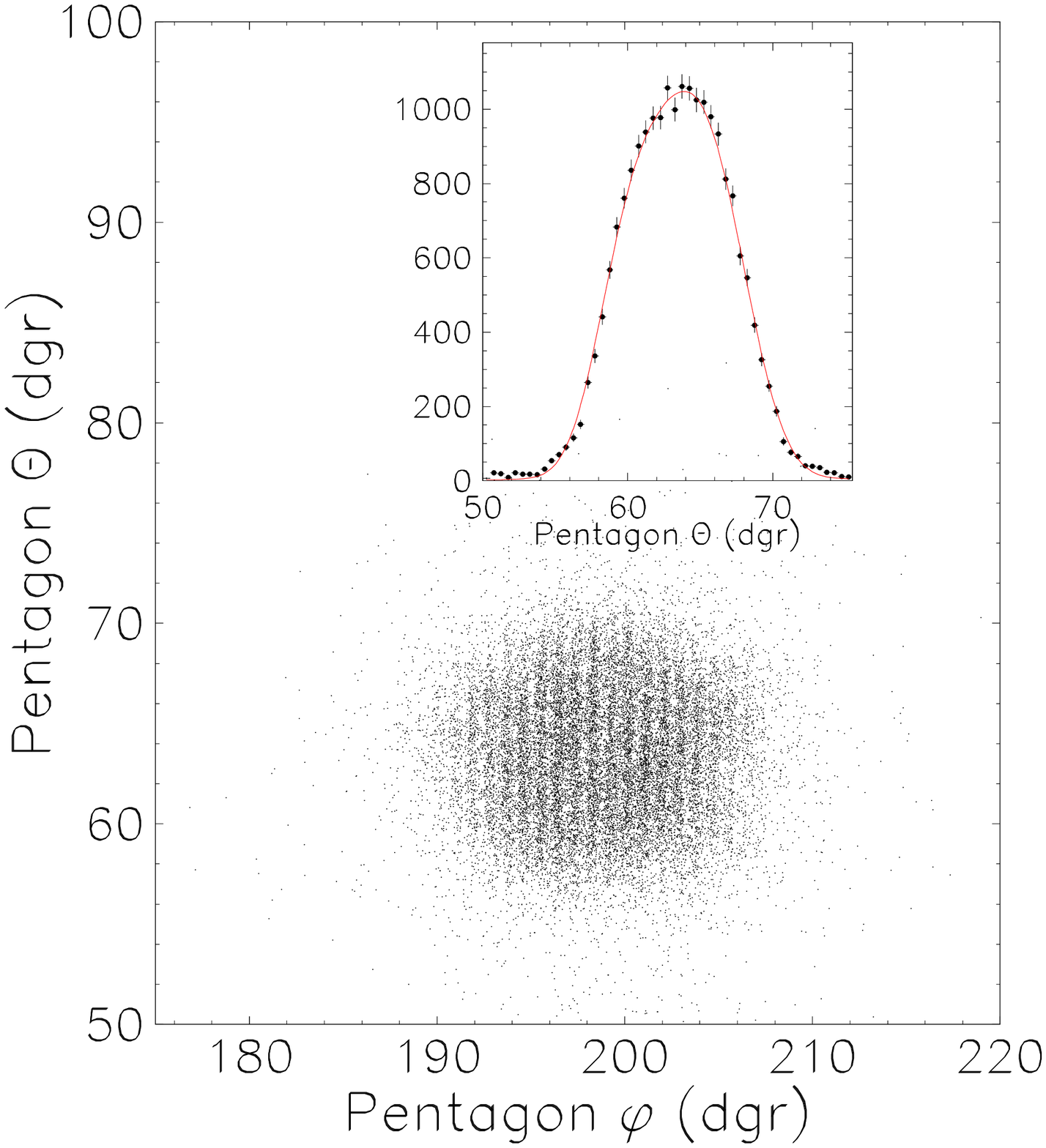,width=2.7in}}
  \vglue 0.0cm
       }\hfill
  \vbox{\hsize=3.1in
  \centerline{\psfig{figure=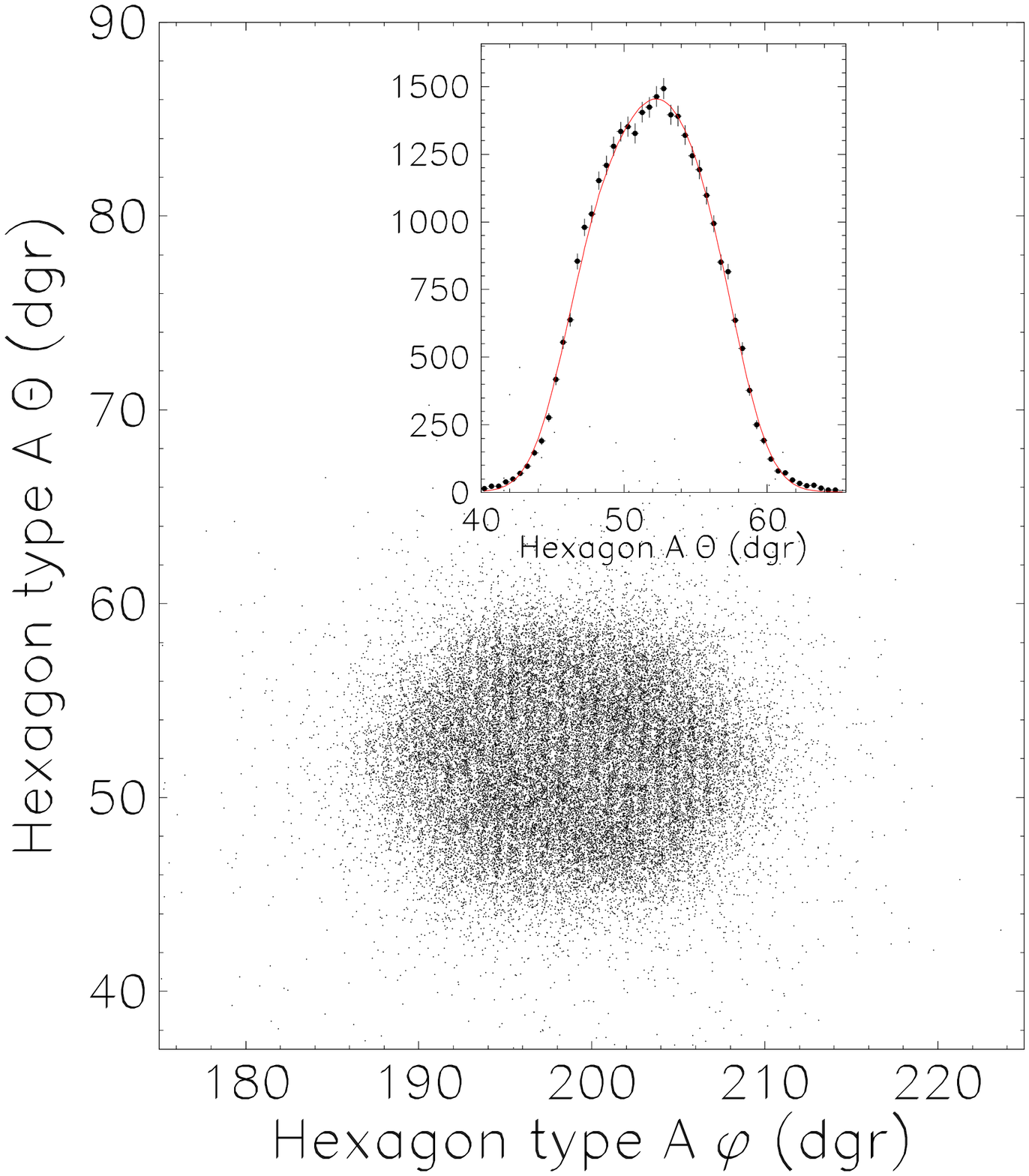,width=2.7in}}
  \vglue 0.0cm
       }
 }
\hbox to \textwidth
 {\vbox{\hsize=3.1in
  \centerline{\psfig{figure=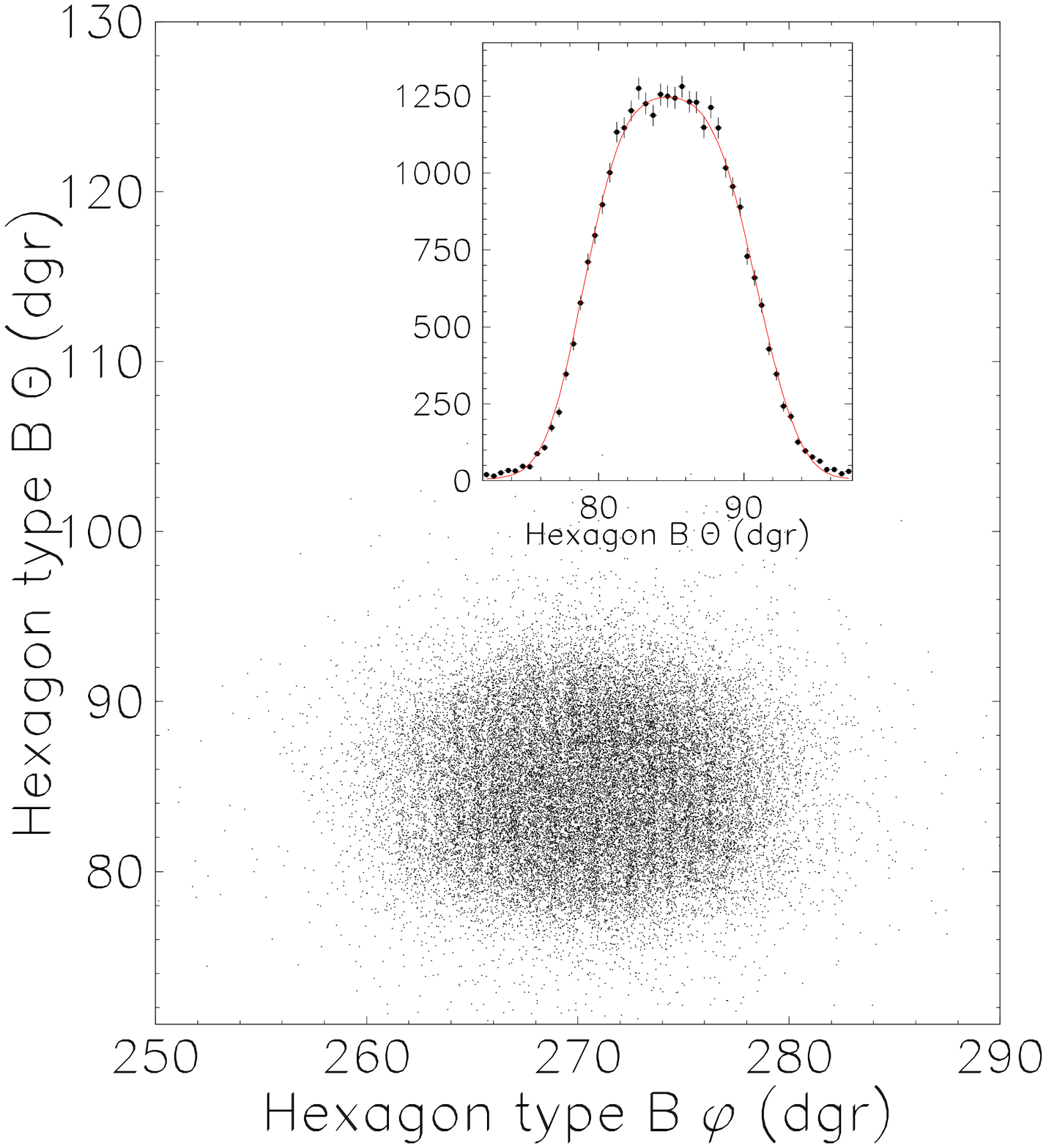,width=2.7in}}
  \vglue 0.0cm
       }\hfill
  \vbox{\hsize=3.1in
  \centerline{\psfig{figure=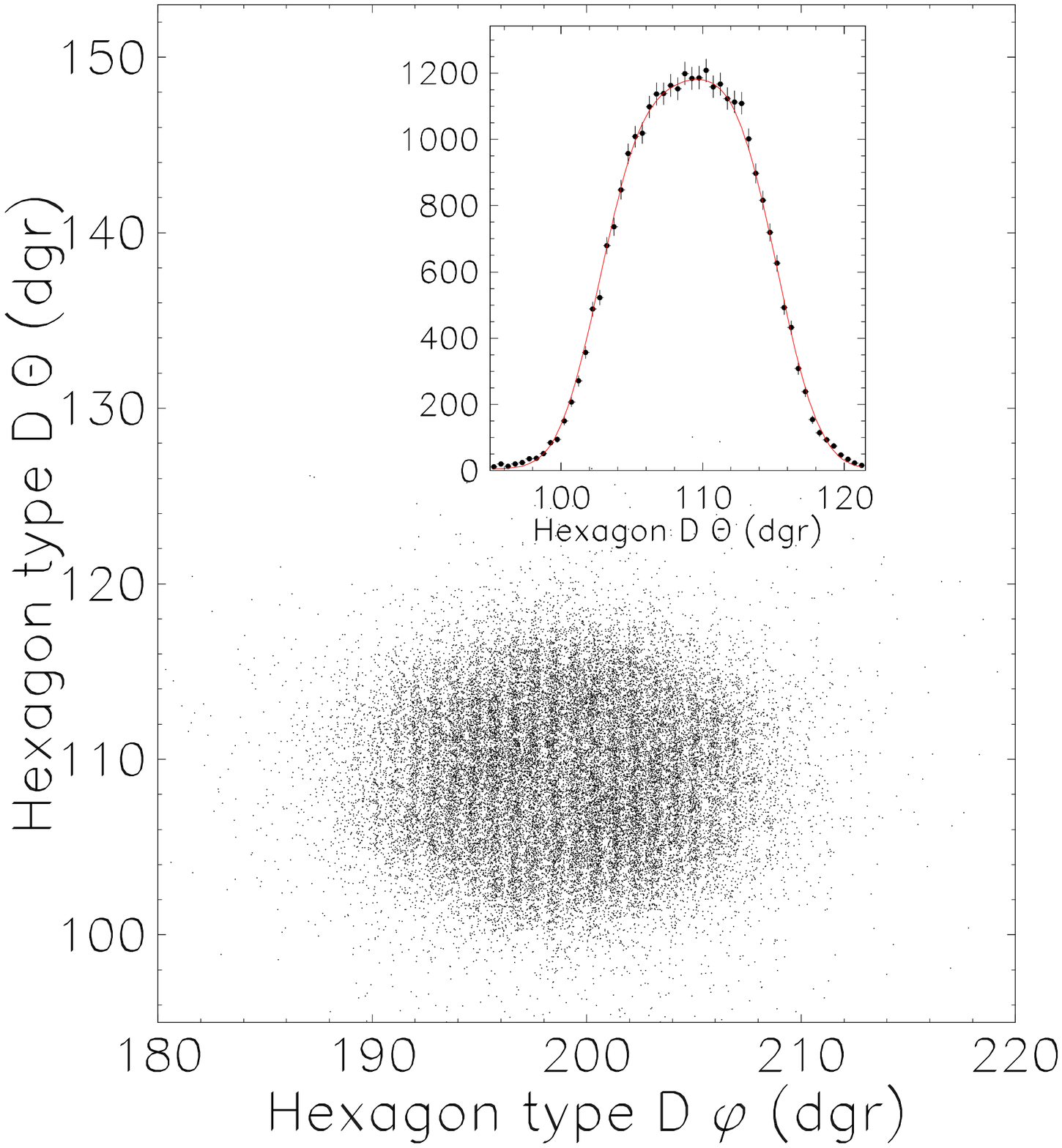,width=2.7in}}
  \vglue 0.0cm
       }
 }
\vglue 1.5cm
\centerline{FIGURE~\ref{fig:pos}}
\vspace*{\stretch{2}}
\clearpage

\vspace*{\stretch{1}}
\centerline{\psfig{figure=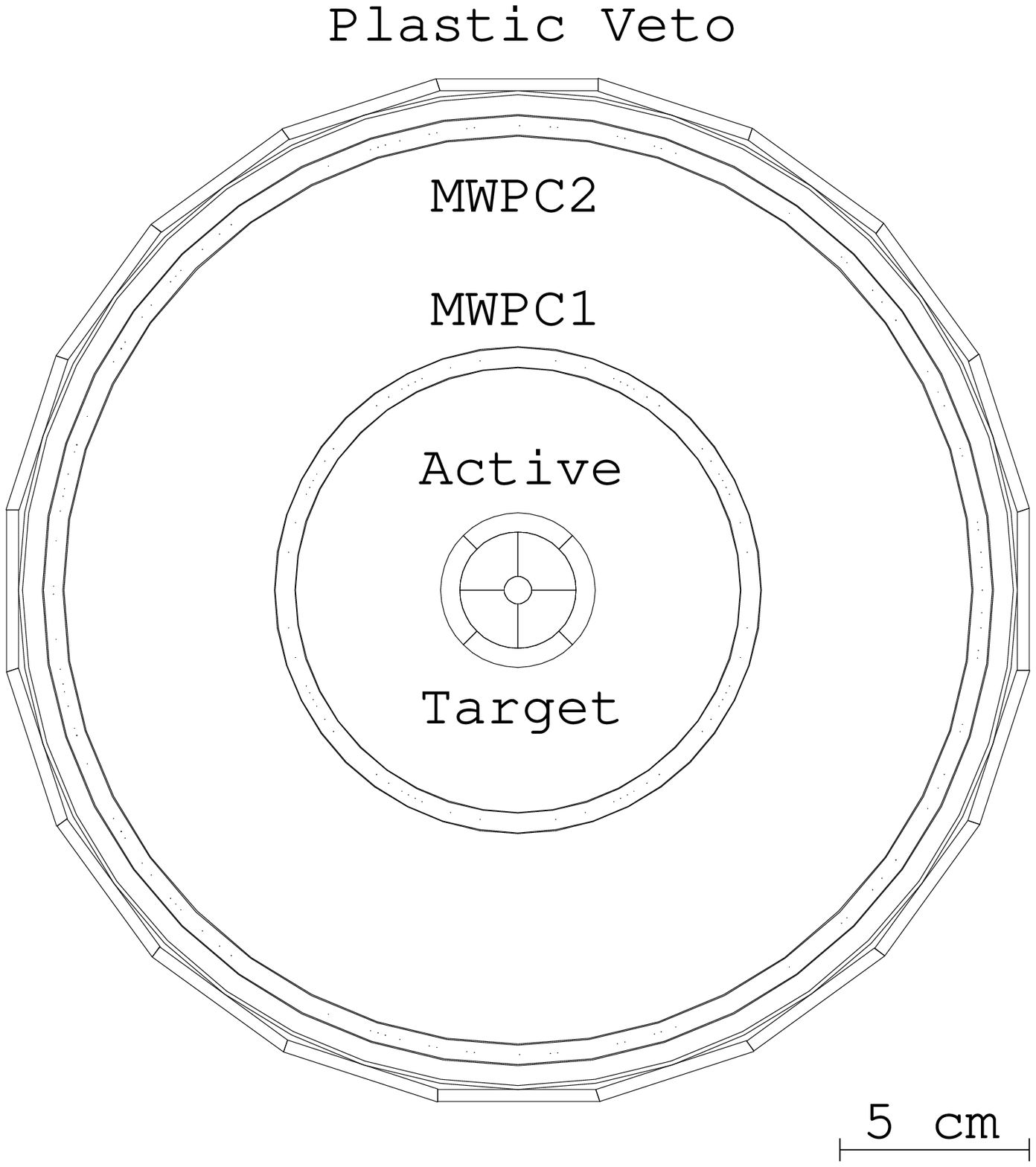,width=12cm}}
\vglue 1cm
\centerline{FIGURE~\ref{fig:central}}
\vspace*{\stretch{2}}
\clearpage

\vspace*{\stretch{1}}
\centerline{\psfig{figure=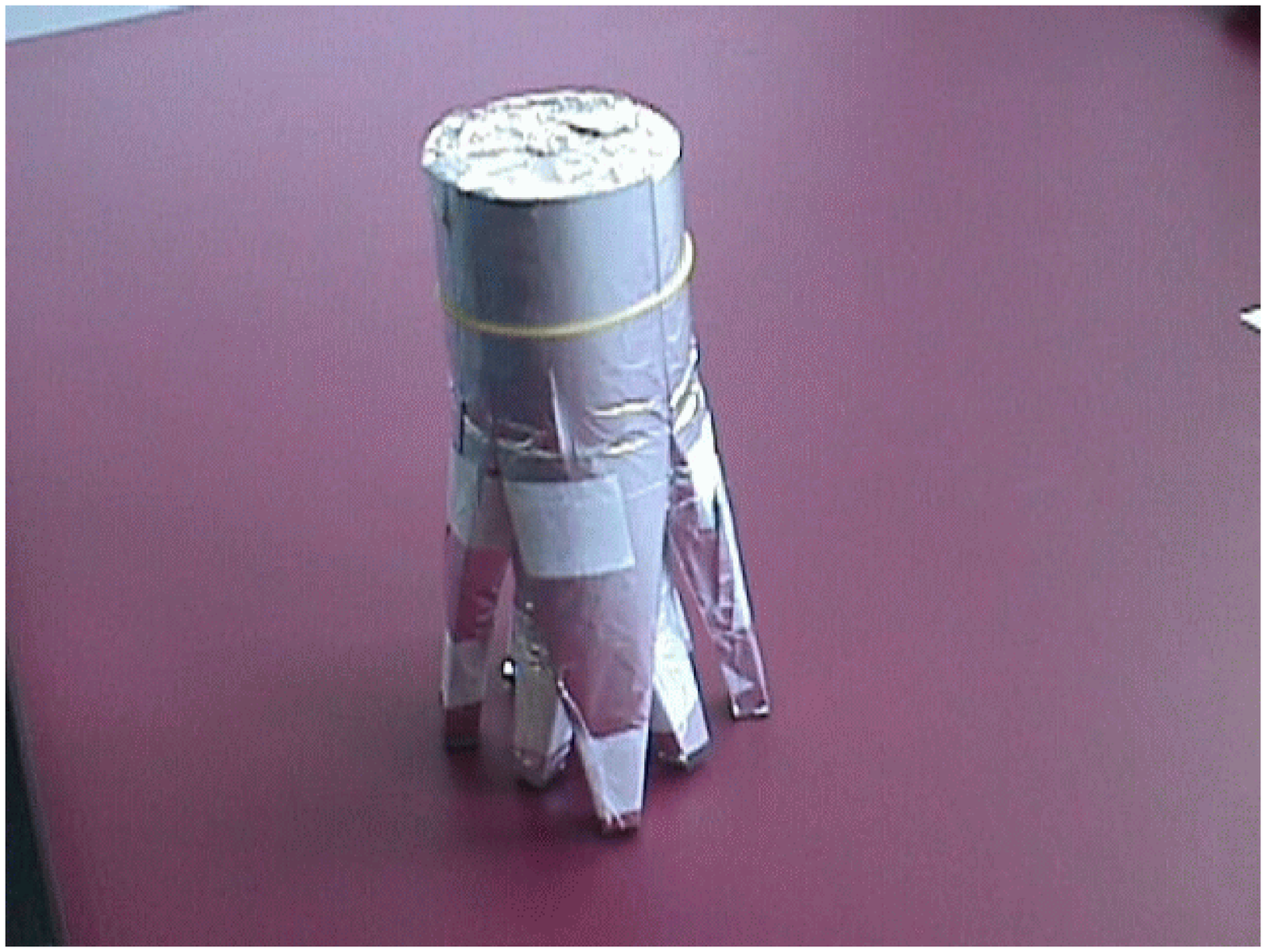,height=10cm}}
\bigskip
\bigskip
\bigskip
\centerline{FIGURE~\ref{fig:photo}}
\vspace*{\stretch{2}}
\clearpage

\vspace*{\stretch{1}}
\centerline{\psfig{figure=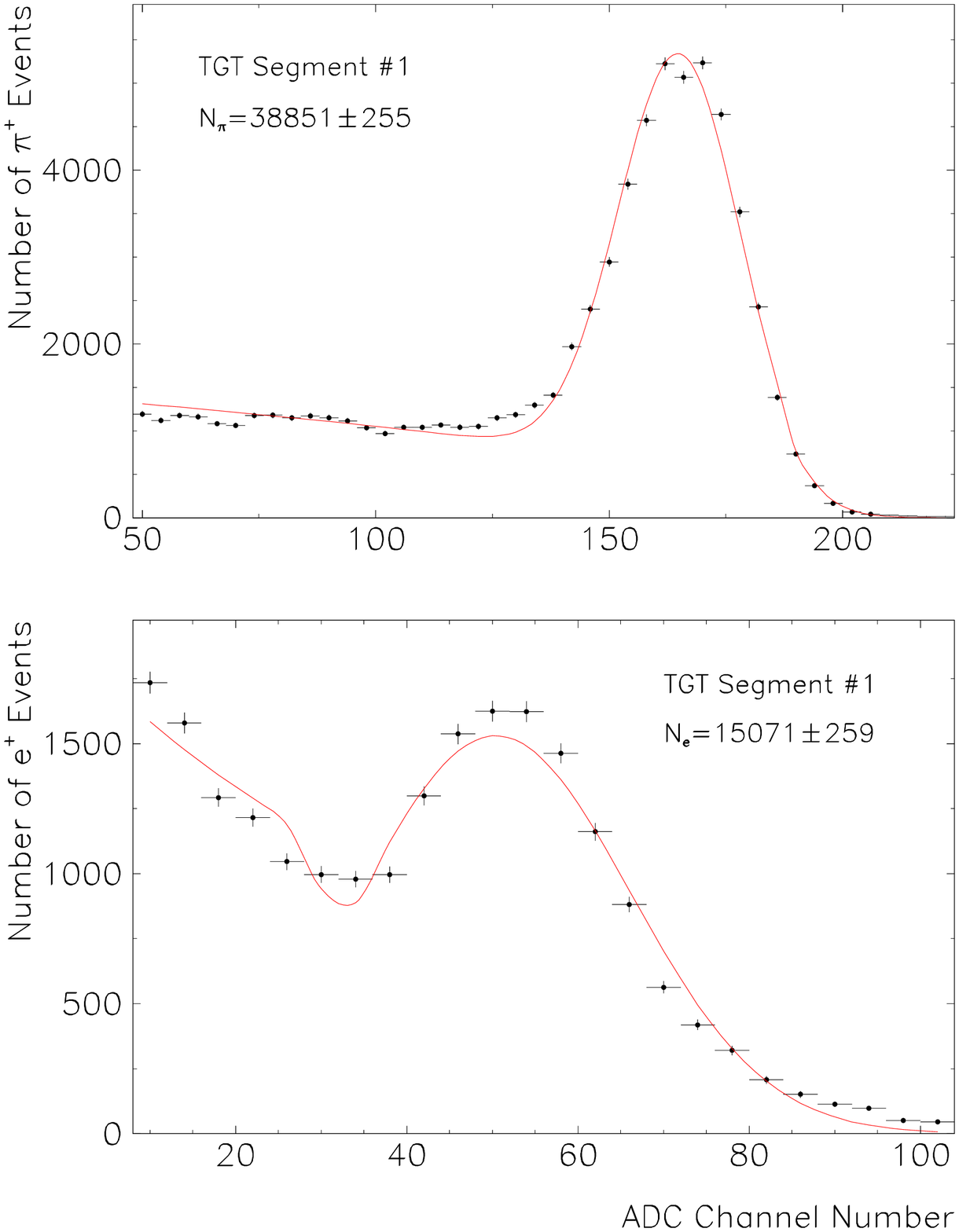,height=20cm}}
\centerline{FIGURE~\ref{fig:pi_e_tgt}}
\vspace*{\stretch{2}}
\clearpage

\vspace*{\stretch{1}}
\centerline{\psfig{figure=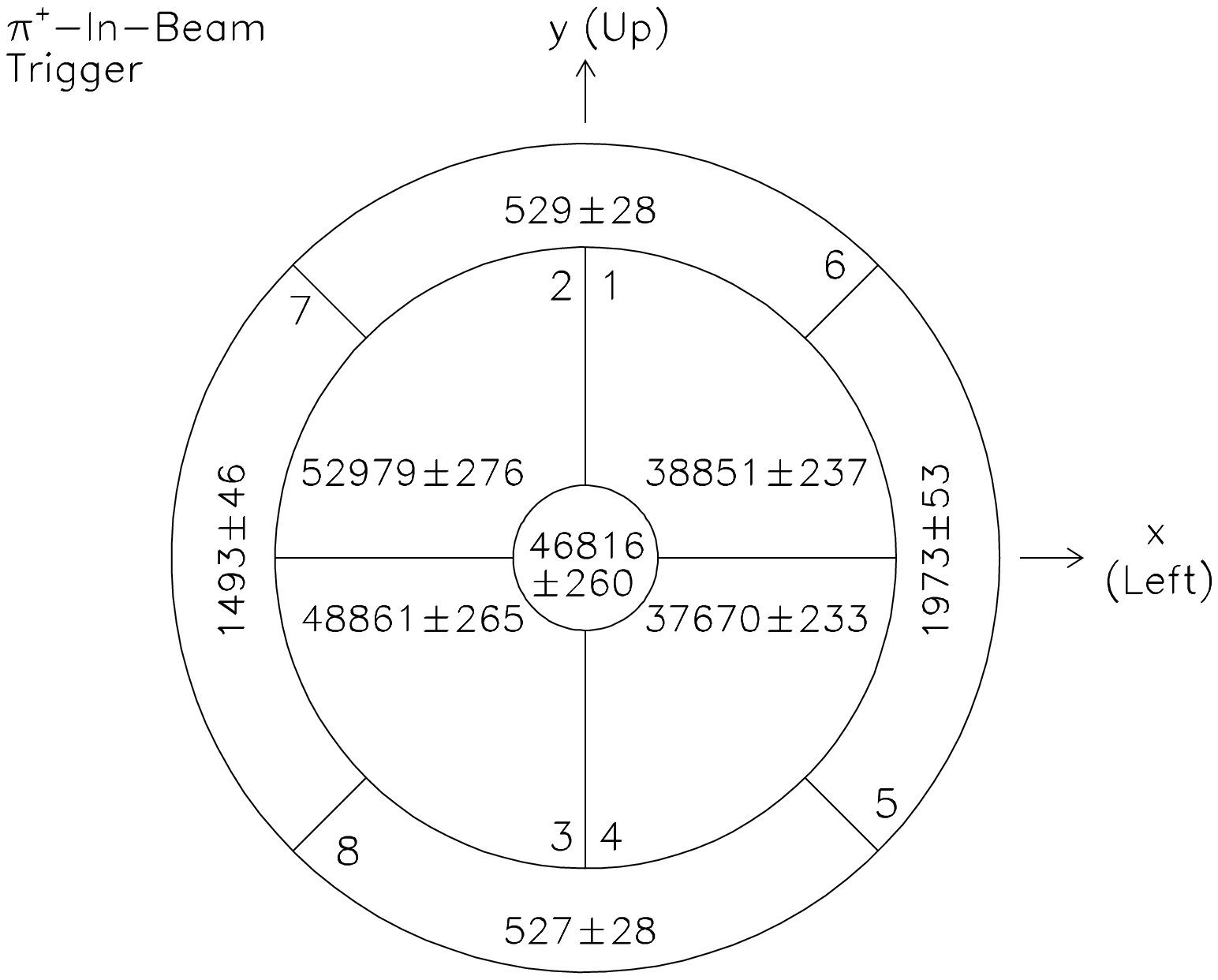,height=12cm}}
\vglue -4.0cm
\centerline{\psfig{figure=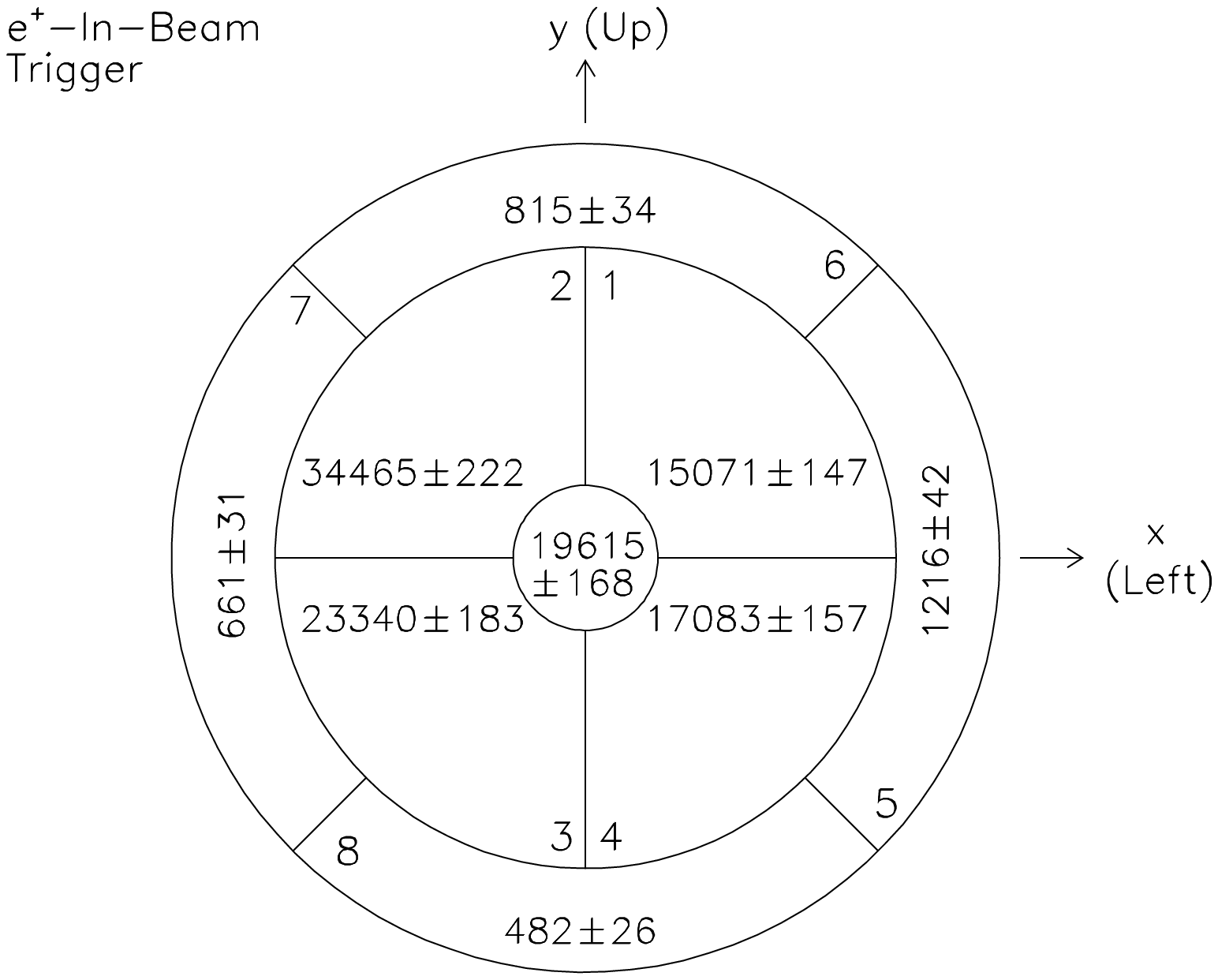,height=12cm}}
\vglue -1.5cm
\centerline{FIGURE~\ref{fig:tgt9}}
\vspace*{\stretch{2}}
\clearpage

\vspace*{\stretch{1}}
\centerline{\psfig{figure=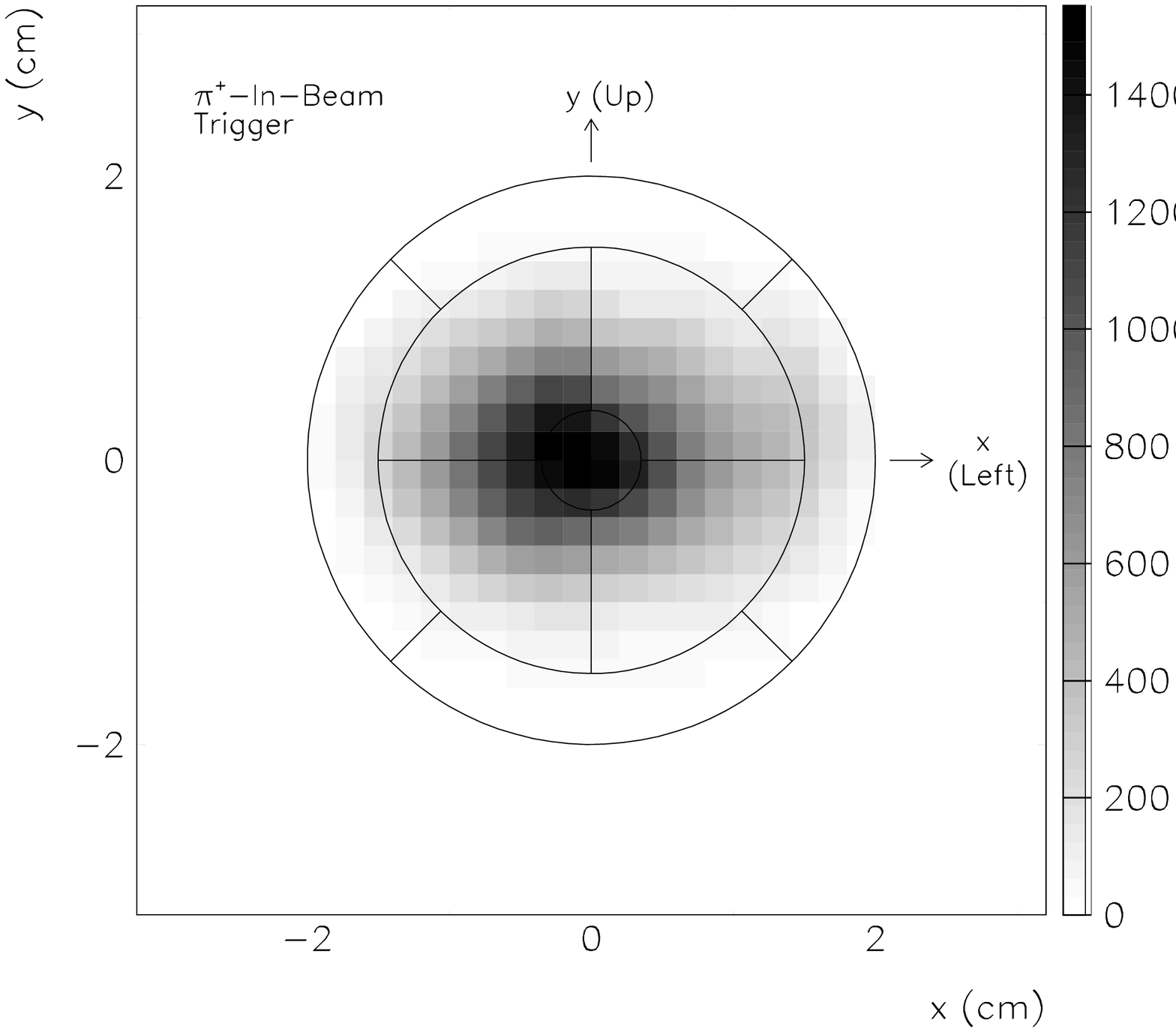,height=14cm}}
\centerline{FIGURE~\ref{fig:extr_2d}}
\vspace*{\stretch{2}}
\clearpage



\vspace*{\stretch{1}}
\centerline{\psfig{figure=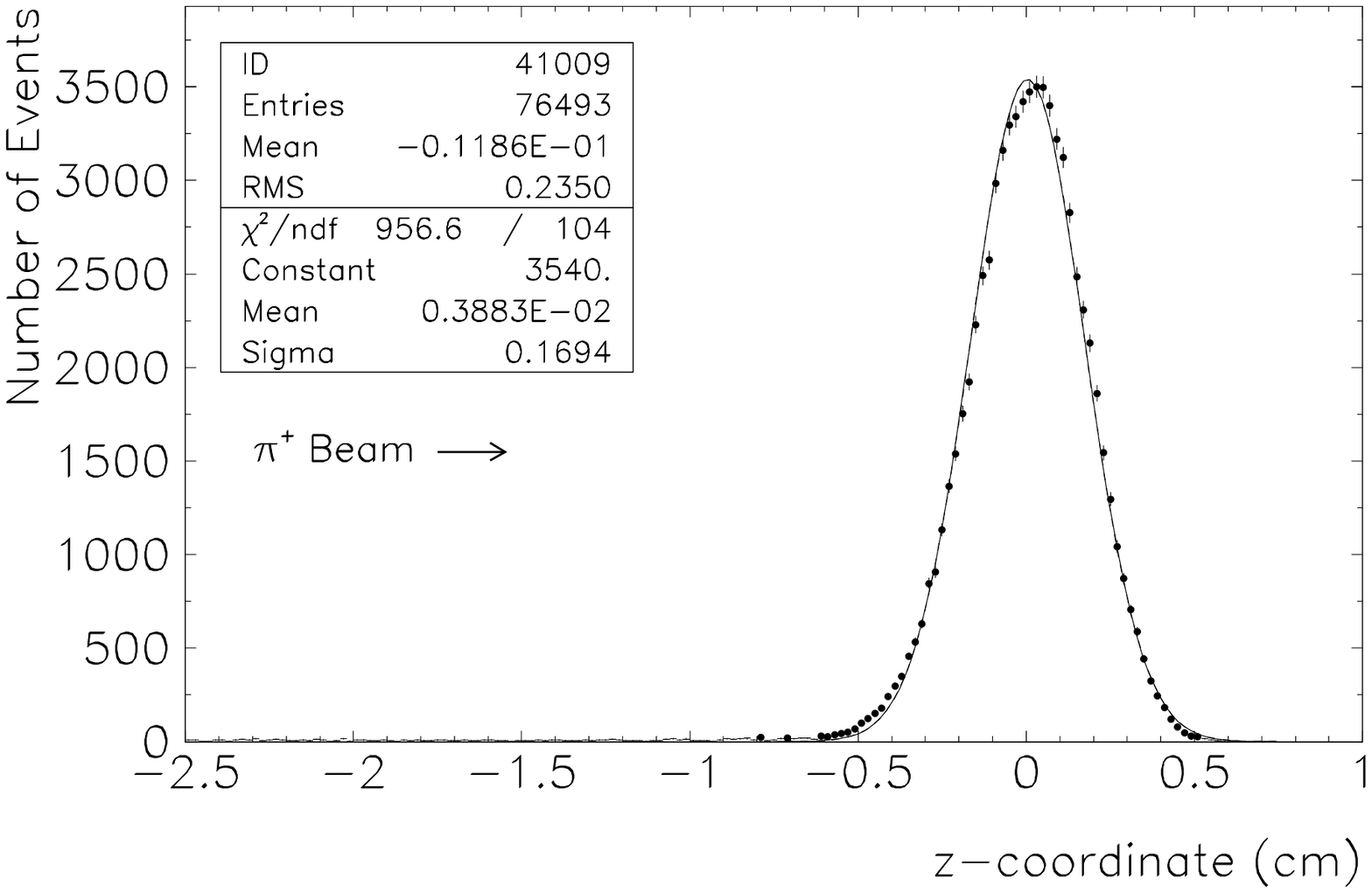,height=22cm}}
\vglue -9.5cm
\centerline{FIGURE~\ref{fig:pistop_z}}
\vspace*{\stretch{2}}
\clearpage

\vspace*{\stretch{1}}
\centerline{\psfig{figure=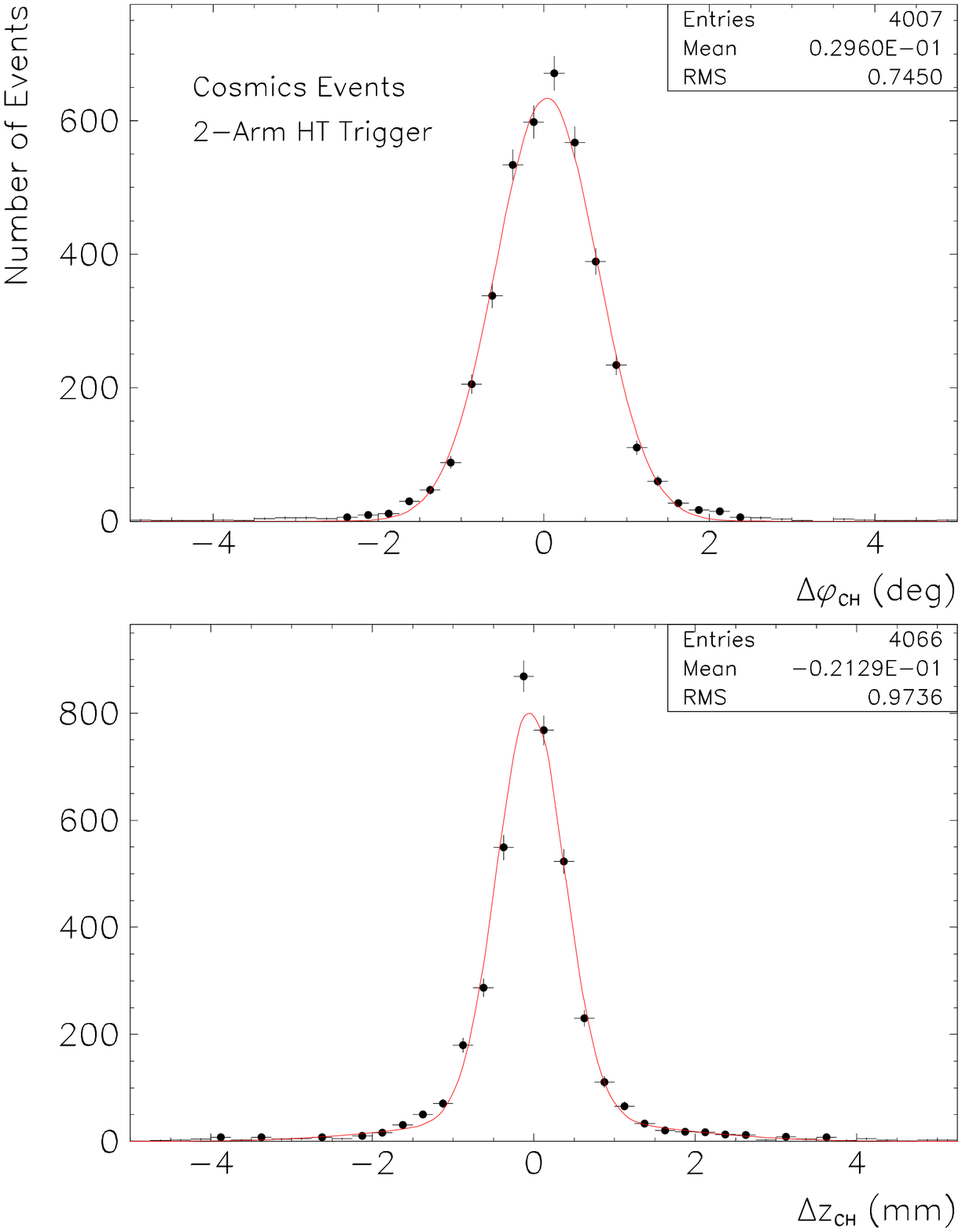,height=21cm}}
\centerline{FIGURE~\ref{fig:mwpc_r}}
\vspace*{\stretch{2}}
\clearpage

\vspace*{\stretch{1}}
\centerline{\psfig{figure=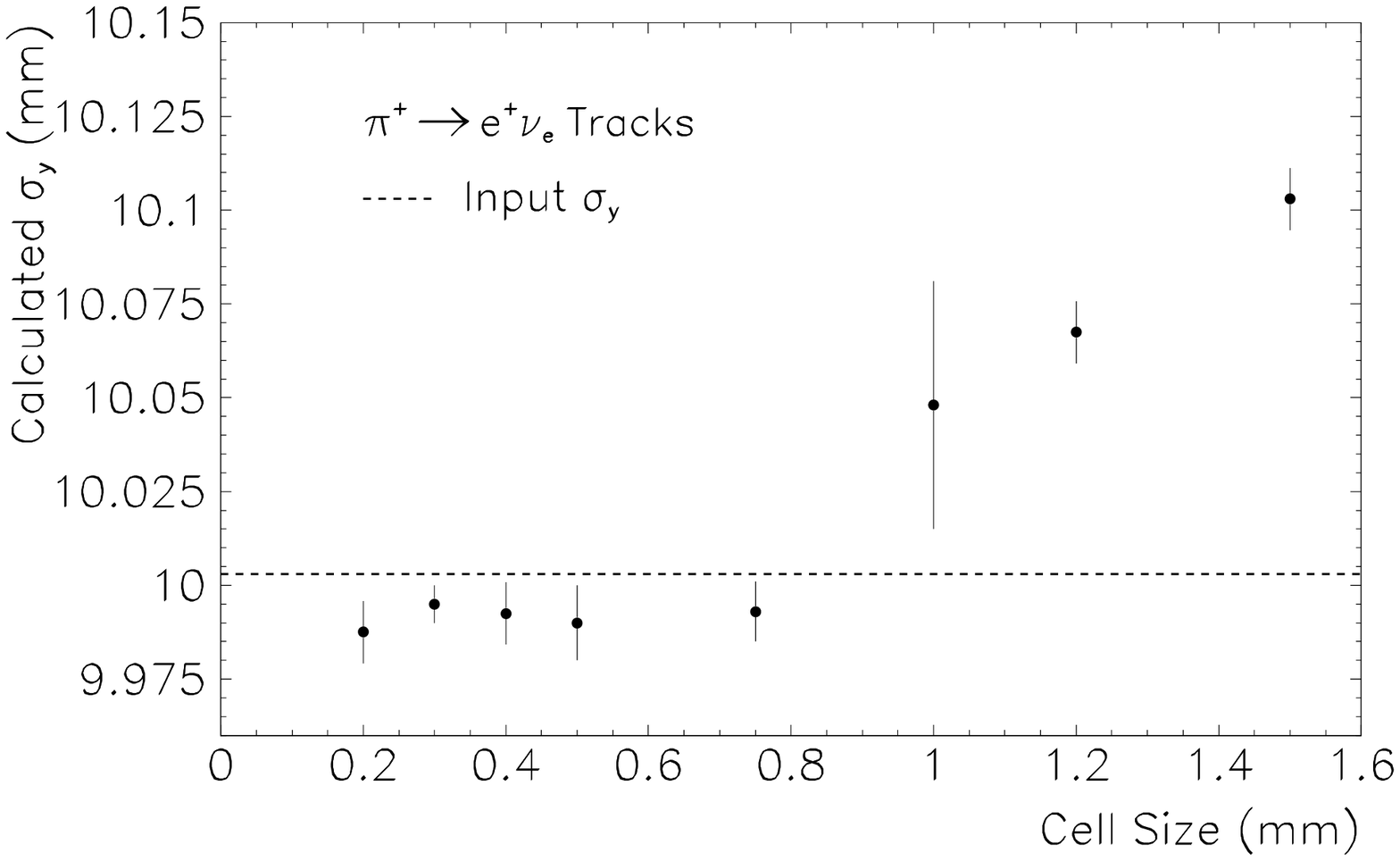,height=21cm}}
\vglue -9cm
\centerline{FIGURE~\ref{fig:tomosys}}
\vspace*{\stretch{2}}
\clearpage

\vspace*{\stretch{1}}
\centerline{\psfig{figure=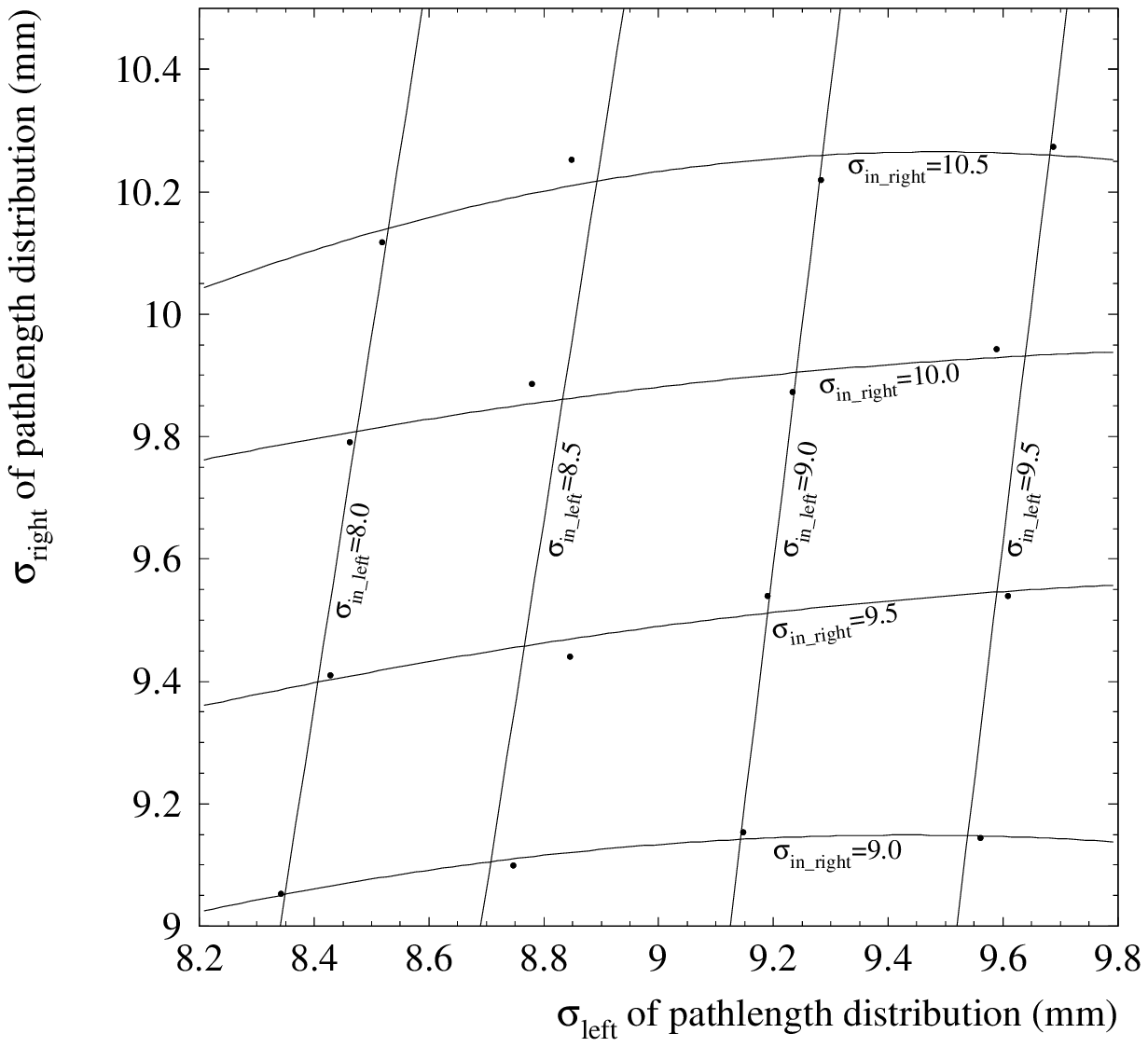,width=18cm}}
\centerline{FIGURE~\ref{fig:lookup}}
\vspace*{\stretch{2}}
\clearpage


\vspace*{\stretch{1}}
\vglue -1.5cm
\centerline{\psfig{figure=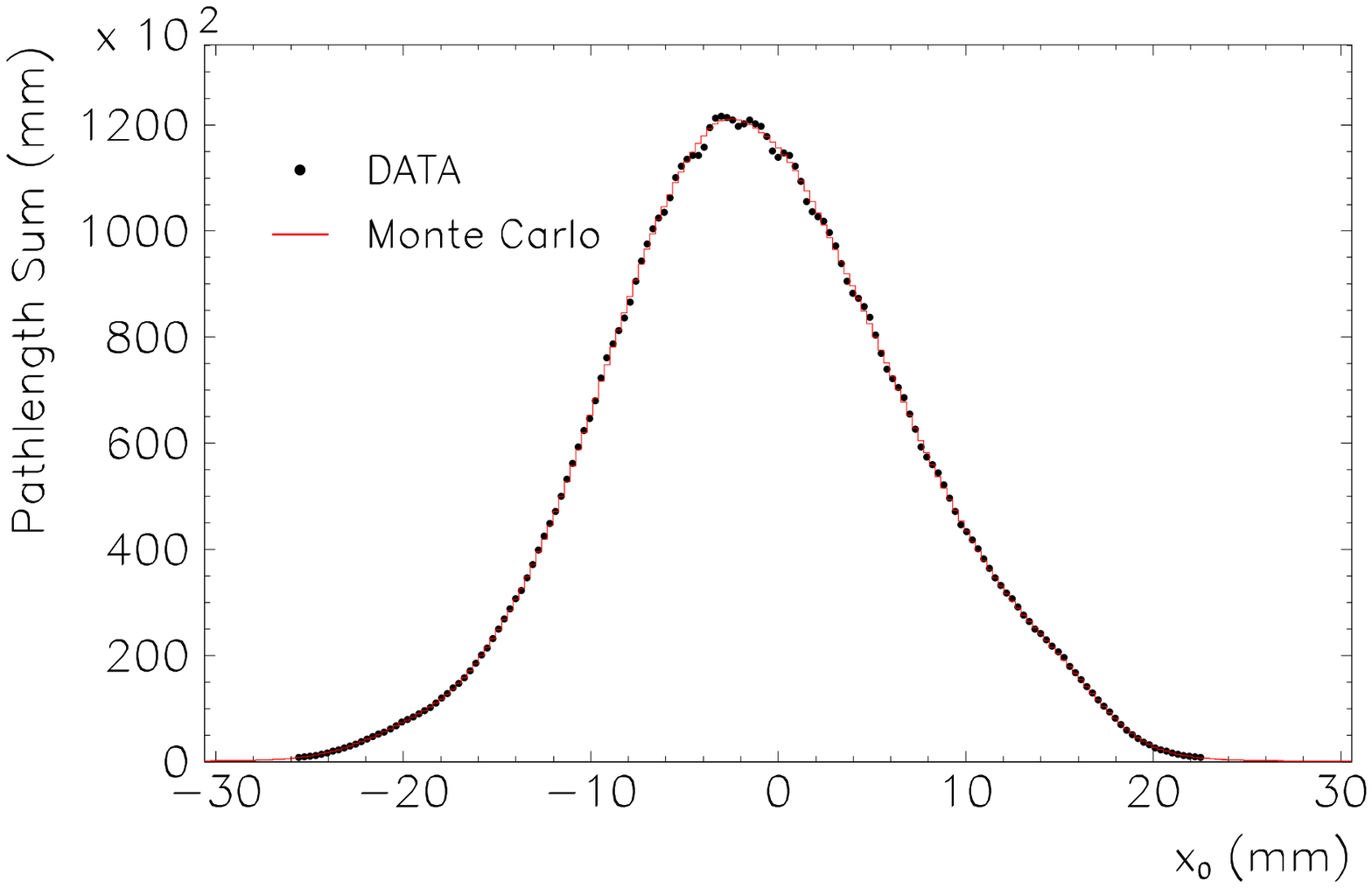,width=12cm}}
\vglue -8.0cm
\centerline{\psfig{figure=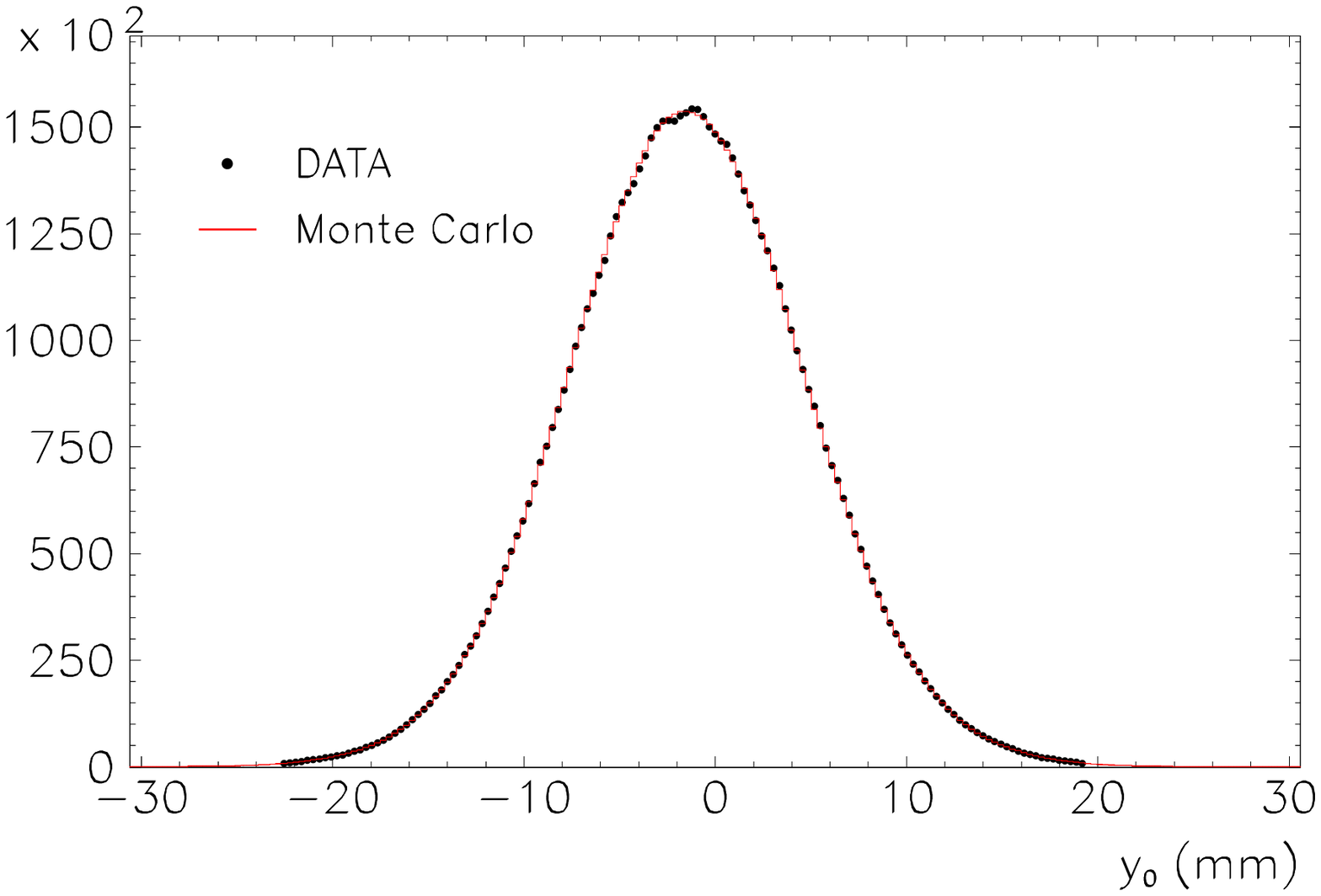,width=12cm}}
\vglue -8.0cm
\centerline{\psfig{figure=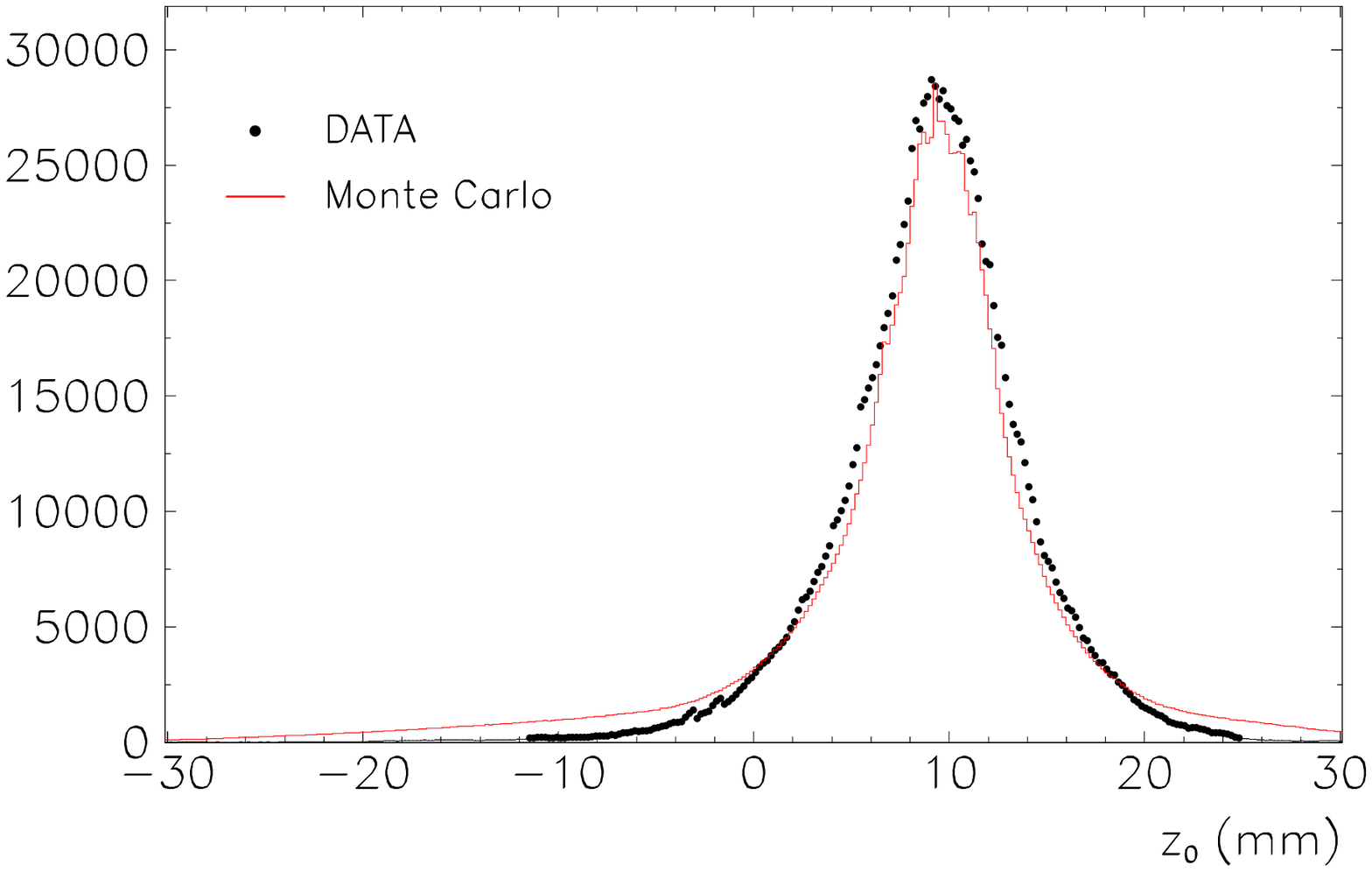,width=12cm}}
\vglue -8.0cm
\bigskip
\centerline{FIGURE~\ref{fig:tomo}}
\vspace*{\stretch{2}}
\clearpage

\vspace*{\stretch{1}}
\vglue -1.5cm
\centerline{\psfig{figure=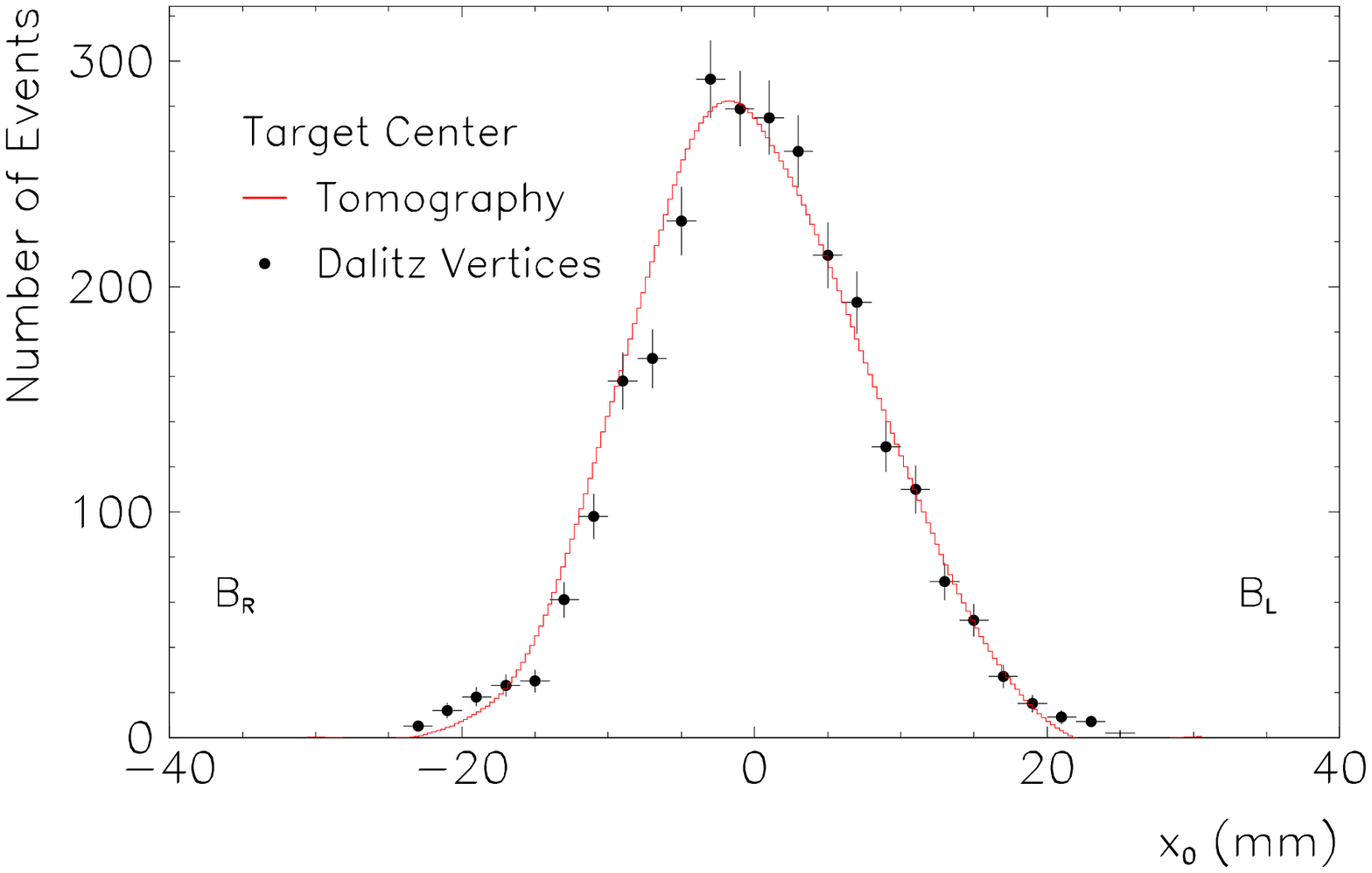,width=12cm}}
\vglue -8.0cm
\centerline{\psfig{figure=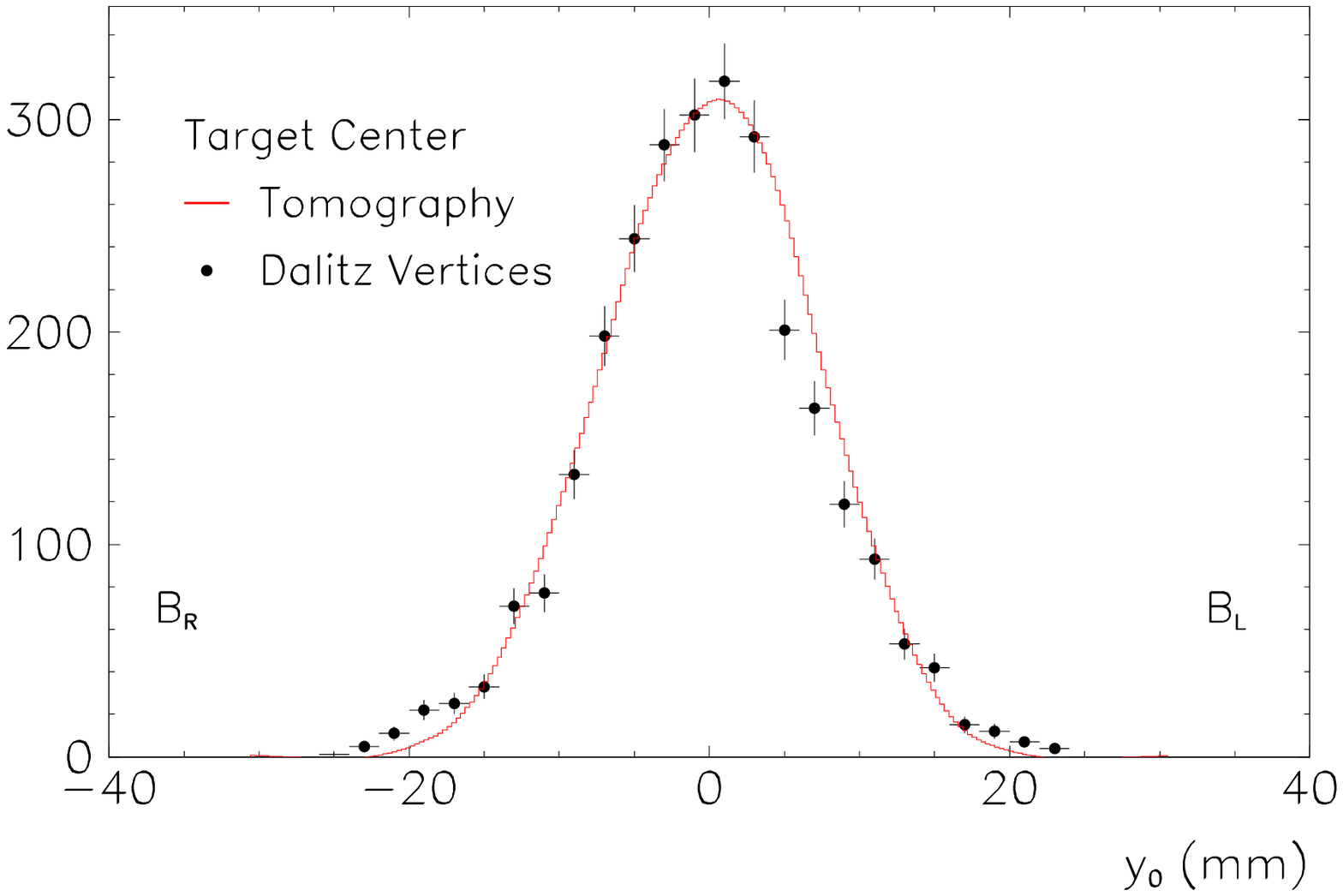,width=12cm}}
\vglue -8.0cm
\centerline{\psfig{figure=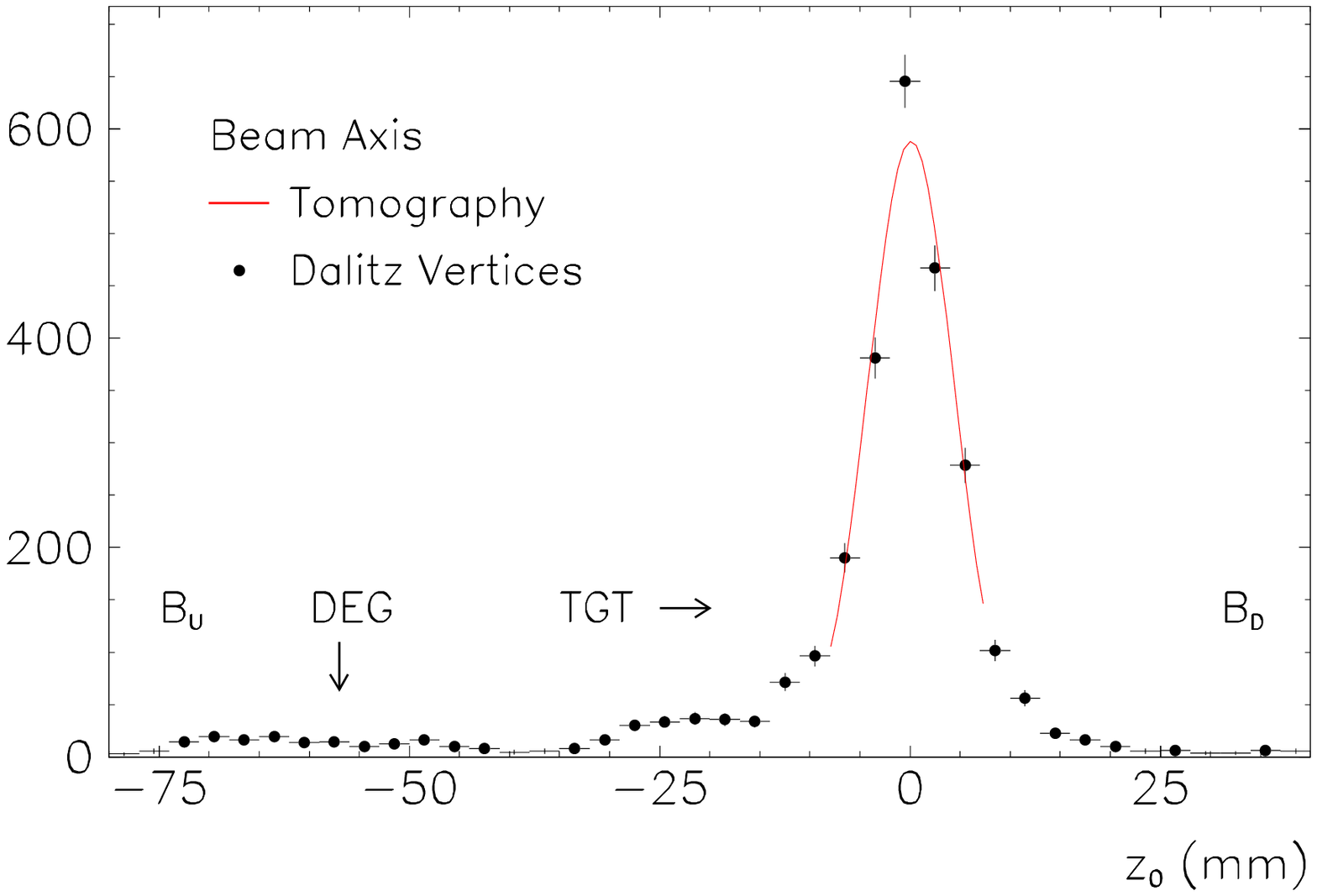,width=12cm}}
\vglue -8.0cm
\bigskip
\centerline{FIGURE~\ref{fig:dal}}
\vspace*{\stretch{2}}
\clearpage


\vspace*{\stretch{1}}
\vglue -1cm
\centerline{\psfig{figure=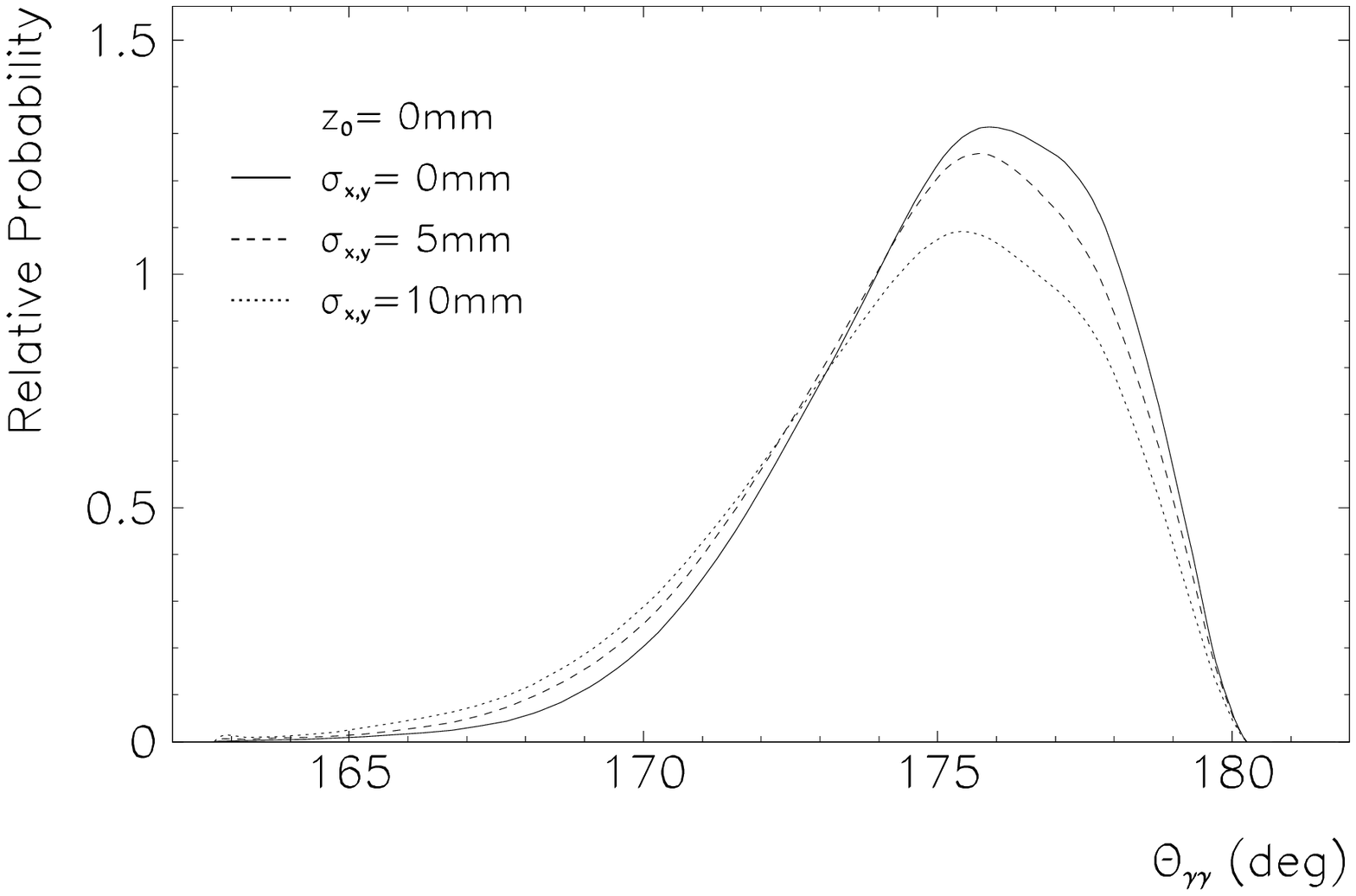,height=7.5cm}}
\centerline{\psfig{figure=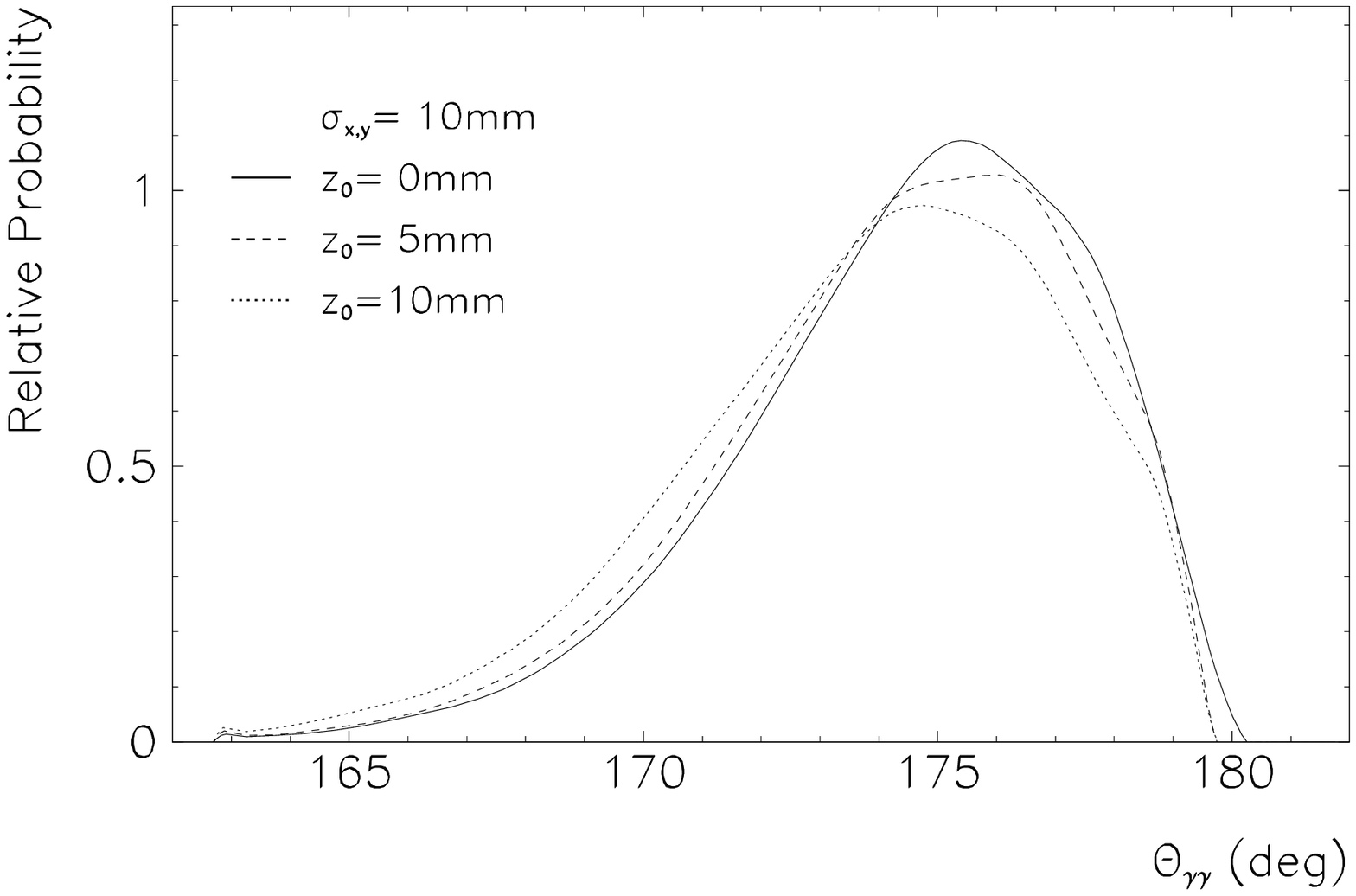,height=7.5cm}}
\centerline{\psfig{figure=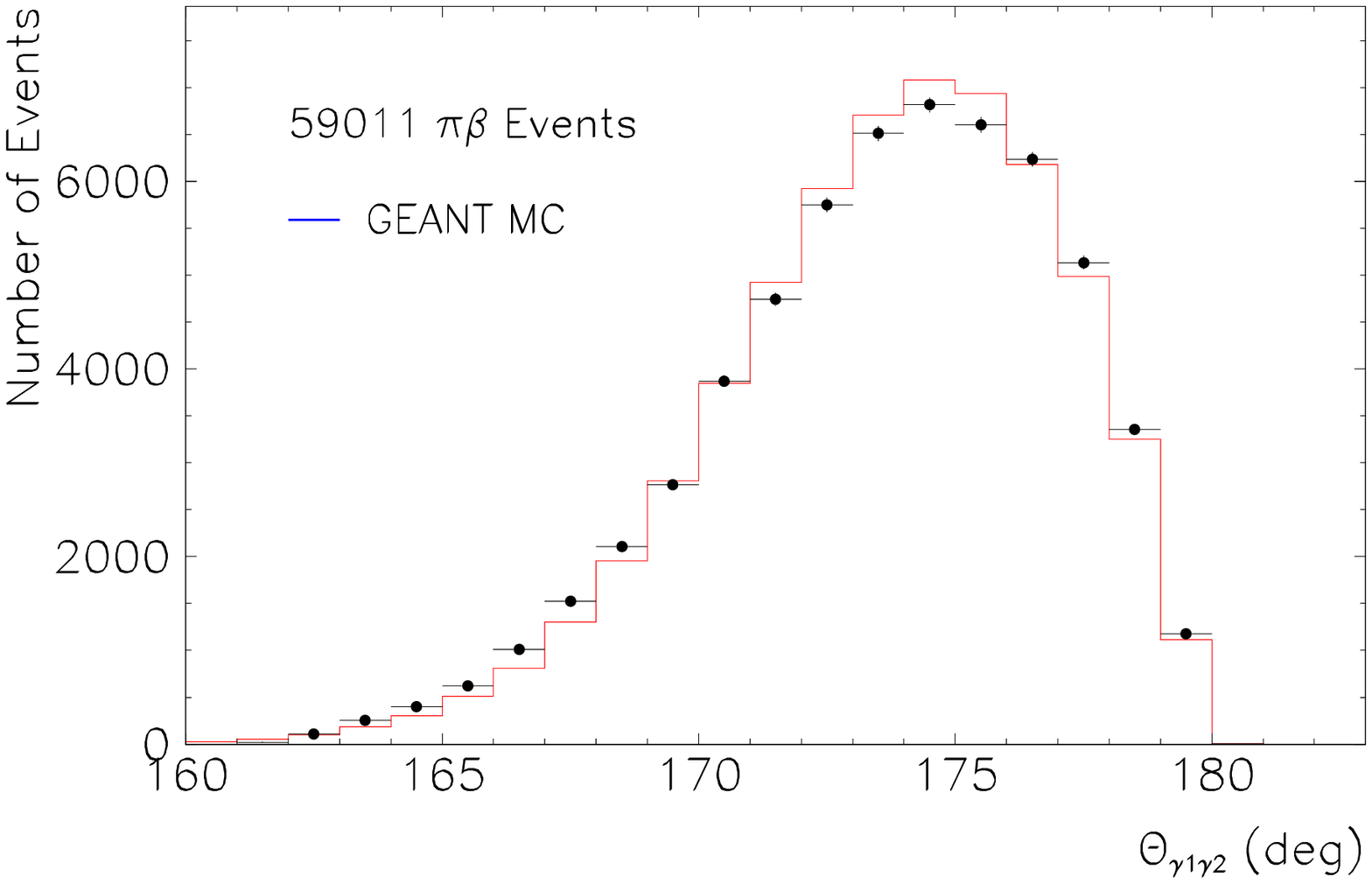,height=7.5cm}}
\medskip
\centerline{FIGURE~\ref{fig:th_gg}}
\vspace*{\stretch{2}}
\clearpage

\vspace*{\stretch{1}}
\centerline{\psfig{figure=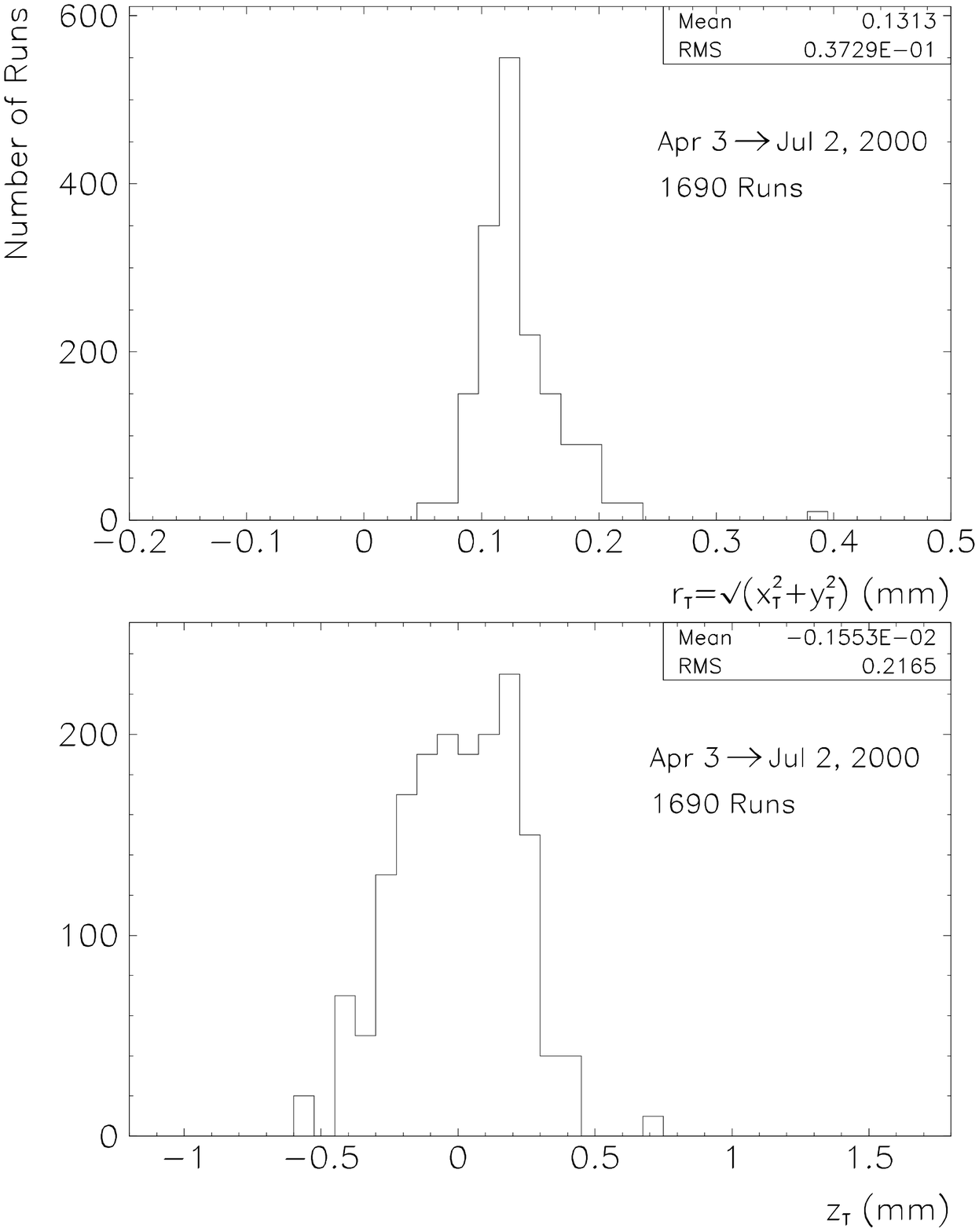,height=20cm}}
\centerline{FIGURE~\ref{fig:beam_stab}}
\vspace*{\stretch{2}}
\clearpage


\clearpage

\begin{thebibliography}{9}

\bibitem{Poc91} 
D. Po\v cani\'c et al., A Proposal for a Precise Measurement of
the $\pi^+\to \pi^0 e^+\nu$ Decay Rate (following PSI R-89.01)
(Paul Scherrer Institute, Villigen, 1991).

\bibitem{Poc04}
D.~Po\v{c}ani\'c et al., Phys. Rev. Lett. 93 (2004) 181803-1-4.

\bibitem{Frl03}
E.~Frle\v{z} et al., Nucl. Instr. and Meth. A 526 (2004) 300.

\bibitem{Psi94}
PSI Users' Guide: Accelerators Facilities (Paul Scherrer 
Institute, Villigen PSI, 1994). 

\bibitem{Bro73}
K.~L.~Brown, D.~C.~Carey, Ch.~Iselin, and F.~Rothacker, Transport, 
a Computer Program for Designing Charged Particle Beam Transport Systems,
CERN Yellow Reports 73-16/80-04 (CERN, Geneva, 1973/1980).

\bibitem{Bro74}
K.~L.~Brown, Ch.~Iselin, and D.~C.~Carey, Decay Turtle, CERN Yellow Report 74-2 
(CERN, Geneva, 1974). 

\bibitem{max}
M.~Bychkov, {\sl Measurements of the Wire Chamber's Displacement},
accessible at {\tt http://\-pibeta.\-phys.\-virginia.\-edu}.

\bibitem{Bru94} 
R.~Brun, F.~Bruyant, M.~Maire, A.~C.~McPherson,
and P.~Zanarini, GEANT 3.21 DD/EE/94-1 (CERN, Geneva, 1994).

\bibitem{Jam89} 
F. James and M. Roos, MINUIT---Function 
Minimization and Error Analysis, CERNLIB Long Write-up D506 
(CERN, Geneva, 1989).

\bibitem{PDG}
K.~Hagiwara~et al., Phys. Rev. D 66 (2002) 010001.
Latest {\tt WWW} update accessible at {\tt http://www-pdg.lbl.gov}.

\bibitem{Kar98}
V.~V.~Karpukhin, I.~V.~Kisel, A.~S.~Korenchenko, S.~M.~Korenchenko, N.~P.~Kravchuk, 
N.~A.~Kuchinsky, N.~V.~Khomutov, and S.~Ritt, Nucl. Inst. and Meth. A 418
(1998) 306.

\bibitem{Gor74}
R.~Gordon, IEEE Trans. Nucl. Sci. NS-21 (1974) 78.

\bibitem{Li04}
W. Li, {\sl A Precise Measurement of the $\pi^+\to \pi^0 e^+\nu$
Branching Ratio}, Ph. D. Thesis (University of Virginia,
Charlottesville, 2004).

\bibitem{Frl97}
E. Frle\v{z}, Complete GEANT3 Description of the PIBETA Detector, 
ac\-ces\-si\-ble at URL 
{\tt ftp:\-//\-\-pi\-be\-ta.\-phys.\-Vir\-gin\-ia.\-EDU} under
{\tt /pub\-/pibeta\-/geant}, (1997).

\bibitem{Kal60}
G.~K\"all\'en, Elementary Particle Physics (Addison-Wesley, Reading, 
1964).

\bibitem{Gin66}
E.~S.~Ginsberg, Phys. Rev. 142 (1966) 1035.

\bibitem{Awe92}
T.~C.~Awes, F.~E.~Obenshain, F.~Plasil, S.~Saini, S.~P.~Sorensen, and
G.~R.~Young, Nucl. Instr. and Meth. A 311 (1992) 130.

\bibitem{Bug86}
L.~Bugge, Nucl. Instr. and Meth. A 242 (1986) 228.



\end{thebibliography}
\end{document}